\documentclass[preprint,10pt]{elsarticle}

\usepackage{graphicx, float}
\usepackage{amssymb}
\usepackage{amsthm}
\usepackage{amsmath}
\usepackage{verbatim}
\usepackage{lineno}
\usepackage{todonotes}
\definecolor{azure}{rgb}{0.0, 0.5, 1.0}
\newcounter{todocounter}

\begin{document}
\begin{frontmatter}
\title{Computational assessment of smooth and rough parameter dependence of statistics in chaotic dynamical systems}

\author[a1,a2]{Adam A. \'Sliwiak\corref{cor}}
\ead{asliwiak@mit.edu}
\author[a1,a3]{Nisha Chandramoorthy}
\ead{nishac@mit.edu}
\author[a1,a2]{Qiqi Wang}
\ead{qiqi@mit.edu}
\address[a1]{Center for Computational Science and Engineering, Massachusetts Institute of Technology (MIT), Cambridge, MA, 02139, USA}
\address[a2]{MIT Department of Aeronautics and Astronautics}
\address[a3]{MIT Department of Mechanical Engineering}
\cortext[cor]{Corresponding author.}
\begin{abstract}
 An assumption of smooth response to small parameter changes, of statistics or long-time averages of a chaotic system, is generally made in the field of sensitivity analysis, and the parametric derivatives of statistical quantities are critically used in science and engineering. In this paper, we propose a numerical procedure to assess the differentiability of statistics with respect to parameters in chaotic systems. We numerically show that the existence of the derivative depends on the Lebesgue-integrability of a certain {\em density gradient} function, which we define as the derivative of logarithmic SRB density along the unstable manifold. We develop a recursive formula for the density gradient that can be efficiently computed along trajectories, and demonstrate its use in determining the differentiability of statistics. Our numerical procedure is illustrated on low-dimensional chaotic systems whose statistics exhibit both smooth and rough regions in parameter space. \end{abstract}

\end{frontmatter}

\section{Introduction}\label{sec:introduction}
Sensitivity analysis is the study of the response of a dynamical system to small perturbations. In this paper, we are interested in the response of long-term or statistical behavior in a chaotic dynamical system, for instance, time-averaged dynamic force coefficients in a turbulent flow \cite{geng-aero}, to small changes in system parameters.
The sensitivities we are interested in are derivatives of long-time averages of observables with respect to parameters. These derivatives are used in gradient-based approaches for design, optimization and control \cite{kirsch-structures,geng-aero,dwyer-turbulence,hu-turbulence, chua-product, weimin-product}, variational data assimilation \cite{ren-assimilation, margulis-assimilation} and uncertainty quantification applications \cite{arriola-book, caicedo-robustness} in various fields of science and engineering.\\ 
In chaotic dynamical systems describing complex phenomena, such as the Earth's climate \cite{lorenz-climate,lea-climate}, molecular transition \cite{hwang-quantum}, and turbulent flow \cite{ni-jfm}, the high sensitivity to perturbations poses a challenge to computing the parametric derivatives of statistics. As a result of this so-called {\em butterfly effect}, the sensitivities of time-averaged quantities with respect to parameters, which can be obtained using traditional methods such as tangent/adjoint equations \cite{jameson-conventional}, automatic differentiation \cite{rackauckas-conventional} and finite-differencing \cite{peter-conventional}, grow exponentially with averaging time window. However, the sensitivities of the statistical or long-time averaged quantities that we are interested in are bounded quantities. In fact, the assumption of {\em linear  response} is generally made, by which the statistics or long-term averages of a chaotic system vary differentiably  with  respect  to  parameters. That is, for small parameter perturbations, we assume that the statistics or long-term averages of observables are linear in the parameter perturbation.
Recently, more sophisticated methods have been proposed to compute the parametric derivative of statistics or linear response. They include shadowing trajectory-based approaches \cite{wang-lssoriginal, ni-nilss}, ensemble averaging \cite{eyink-ensemble}, transfer operator-based \cite{blonigan-pdf} methods, and variational techniques \cite{lasagna-variational}. Other modern approaches are derived based on the Fluctuation-Dissipation Theorem \cite{abramov-fdt}, Ruelle's linear response theory \cite{chandramoorthy-s3, ni-lra}. While these methods are recently being used in some practical chaotic systems, including low-Reynolds number turbulent flows, climate models of intermediate complexity, and so forth \cite{ni-jfm, chandramoorthy-applications, blonigan-ks, bodai-climate}, the underlying assumption of linear response may itself be violated in other systems. In fact, beyond the classic example of the logistic map \cite{may-logistic,blonigan-pdf} non-smooth response, as demonstrated in \cite{chandramoorthy-shadowing}, one-dimensional chaotic systems can be constructed that exhibit arbitrarily large linear responses: an imperceptible parameter perturbation causes a drastic change in statistics.

As shown by Ruelle \cite{ruelle-original}, linear response is rigorously true in uniformly hyperbolic systems, which represent the simplest setting in which chaotic attractors occur. Based on Gallavotti and Cohen \cite{galavotti-hypothesis}, it is popularly believed that many highly dissipative dynamical systems found in nature behave as if they were uniformly hyperbolic. This conjecture, known as the chaotic hypothesis, is supported by numerical computations of sensitivities of statistical quantities in many popular physical models, including PDE models for turbulence governed by Kuramoto-Sivashinsky equation \cite{blonigan-ks} and the 3D Navier-Stokes equation \cite{ni-jfm}. However, various models used in climate modeling and geophysical fluid dynamics indicate the chaotic hypothesis cannot always be applied. For example, a rough statistics-parameter relationship has been observed in an El-Ni\~no Southern Oscillation climate model in \cite{chekroun-rough}; other work describes violations of linear response on climate models around atmospheric blocking events \cite{gritsun-blocking}. Wormell and Gottwald approach the question of existence of linear response from a statistical mechanical perspective \cite{wormell-macroscopic}. They construct a worst-case prototype of a macroscopic system where linear response is upheld despite its failure in the constituent microscopic subsystems. Besides this theoretical insight, the authors caution \cite{wormell-sensitivity} that detecting the failure of linear response na\"ively using the statistics-vs-parameter curve is too data-intensive, and detailed information about the invariant probability distribution is required. Chekroun {\em et al.}'s work \cite{chekroun-rough} provides a potential mathematical as well as a numerically verifiable procedure for the detection of smoothness of parameter dependences. In particular, a relation between the spectral gap of the transfer operator \cite{baladi-book}, and the smoothness of statistical quantities in parameter space is established. In \cite{gritsun-rough}, a generalization of the Fluctuation-Dissipation Theorem has been used to study the response attributes of an atmospheric general circulation model, and to verify the applicability of the linear response theory for that particular model.

The main contribution of this work is an alternative numerical procedure for the detection of bounded linear response. Our approach is based on using detailed statistical information about the underlying probability distribution, which is represented in the {\em density gradient} function (see Section \ref{sec:density-gradient} and \ref{sec:lorenz} for the definition; see also \cite{sliwiak-1d} for an intuitive description of the density gradient function). However, we are able to compute this fundamental function simply from time series information by developing an efficient ergodic-averaging method. We empirically find that when the computed density gradient is Lebesgue-integrable, linear response holds. This observation can also be mathematically corroborated by integration-by-parts on Ruelle's linear response formula. In particular, we build on our previous work in which we regularize Ruelle's formula and describe an algorithm to compute the regularized formula, known as space-split sensitivity or S3. In this work, we utilize the S3 formula to propose a computable criterion for differentiability of statistics. Thus, the numerical assessment of the validity of linear response presented in this paper has two attractive features: i) it is efficiently computable along trajectories, ii) it produces ingredients that lead to the value of the derivative, when linear response holds. 

The main body of this paper is divided into six sections. In Section \ref{sec:onion}, we analyze a representative one-dimensional chaotic map, which we call the onion map, whose statistical quantities exhibit both smooth and non-smooth behavior. A mathematical argument showing the relation between the derivative of statistics and the density gradient function is presented in Section \ref{sec:density-gradient}. In the same section, we derive a recursive formula for the density gradient function using the measure preservation property and provide implementation details. In Section \ref{sec:onion-results}, we numerically analyze the distribution of $g$ and estimate the H\"older exponent of the statistics-parameter relation to validate our results. Based on the numerical results, we propose a computable mathematical criterion for assessing the smoothness of statistics. Section \ref{sec:lorenz} generalizes our conclusions from Sections \ref{sec:density-gradient}-\ref{sec:onion-results} to higher-dimensional systems and demonstrates numerical results using the Lorenz '63 system as an example. Finally, Section \ref{sec:conclusions} concludes this paper. 

\section{Onion map as an example of one-dimensional chaos}\label{sec:onion}
\begin{figure}
\begin{minipage}{0.48\textwidth}
    \centering
    \includegraphics[width=1.\textwidth]{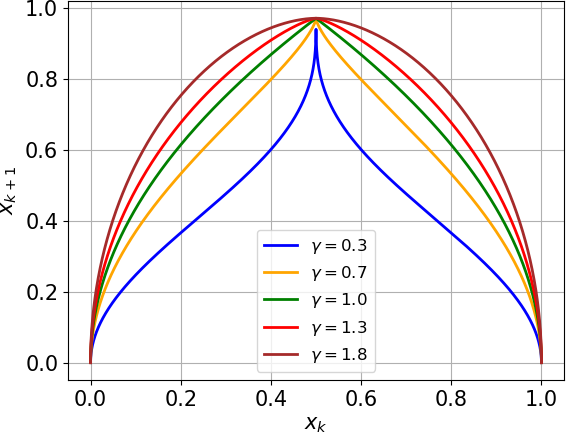}
\end{minipage}
\begin{minipage}{0.51\textwidth}
    \centering
    \includegraphics[width=1.\textwidth]{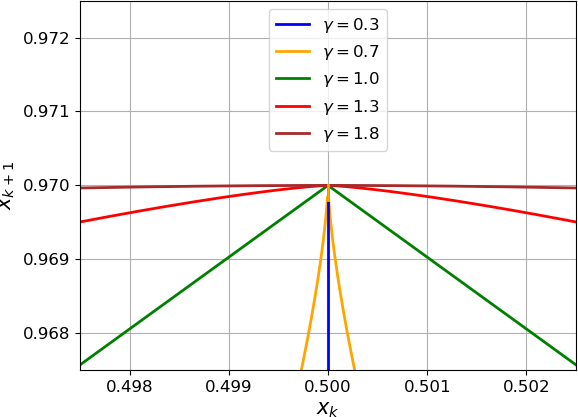}
\end{minipage}
    \caption{Illustration of the onion map at $h = 0.97$ and its dependence on $\gamma$. The right-hand side plot zooms in the region in the vicinity of the tip. }
    \label{fig:onion_map}
\end{figure}

\begin{figure}
\begin{minipage}{0.51\textwidth}
    \centering
    \includegraphics[width=1.\textwidth]{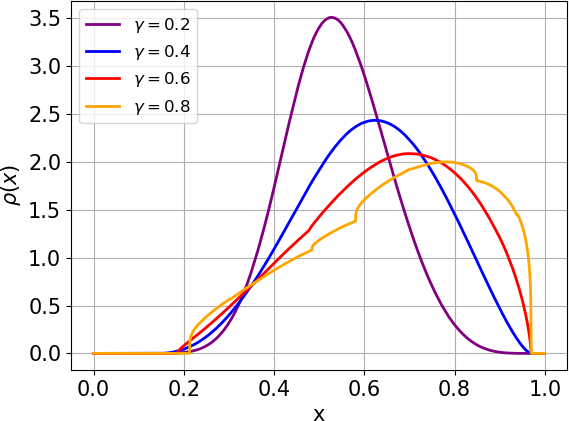}
\end{minipage}
\begin{minipage}{0.49\textwidth}
    \centering
    \includegraphics[width=1.\textwidth]{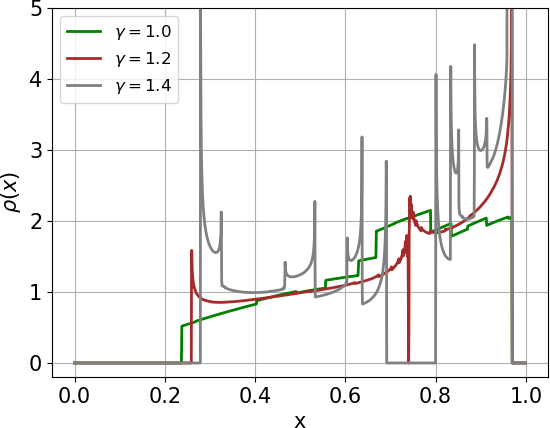}
\end{minipage}
    \caption{Density distribution $\rho(x)$ generated for the onion map (Eq. \ref{eqn:hybrid}) at $h=0.97$. To generate $\rho(x)$, we divided the domain $x\in[0,1]$ into $K = 2048$ bins of equal width, counted the number of times the trajectory passes through each bin. We used $N = 41,943,040,000$ samples to per histogram. The obtained histogram is normalized, through the multiplication by $K/N$, to satisfy the axiom of unit area.}
    \label{fig:hybrid_density}
\end{figure}

At the beginning of our discussion, we present a simple chaotic map that features both smooth and non-smooth (rough) behavior. In that map, the degree of smoothness of statistical quantities strictly depends on the value of its parameters.  Let us consider a one-dimensional map, $\varphi:[0,1]\to[0,h]$, defined as follows,
\begin{equation}
\label{eqn:hybrid}
    x_{k+1} = \varphi(x_k;\gamma,h) =
   h\sqrt{1-|1-2x_{k}|^{\gamma}}.
\end{equation}
where $\gamma>0$, $1\geq h>0$ are map parameters. To simplify the notation, we will skip the parameters in the argument list, i.e., $\varphi(x_k)=\varphi(x_k;\gamma,h)$. Figure \ref{fig:onion_map} depicts Eq. \ref{eqn:hybrid} at some selected parameter values. 
Due to its characteristic shape, this map will be further referred to as the onion map. While the proportionality parameter $h$ only affects the range of $\varphi$, the exponent $\gamma$ has a significant impact on the function shape in the vicinity of the tip located at $(0.5,h)$. If $\gamma<1$, the tip is sharp, i.e., the derivative $\varphi'(0.5)$ does not exist, and its shape resembles the cusp map, as defined and illustrated in \cite{blonigan-pdf}. The cusp map has been used in modeling as a one-dimensional simplification of the Lorenz '63 system \cite{mehta-cusp}. We observe the tip blunts when $\gamma$ gets larger than 1, and the shape of $\varphi$ converges to the well-known logistic map \cite{may-logistic} as $\gamma$ approaches the value of 2. Given the cusp map and logistic map feature smooth and non-smooth statistical behaviors \cite{blonigan-pdf}, the onion map, which combines both of them, is a perfect example of a map with varying regularity of statistical quantities. In our numerical examples, we fix the value of $h$ to $0.97$ and consider different values of $\gamma$.\\
One can easily verify that the only Lyapunov exponent (LE) $\lambda$, defined as
\begin{equation}
    \label{eqn:lyapunov_exponent}
    \lambda = \lim_{N\to\infty}\frac{1}{N}\sum_{k=0}^{N-1}\log\left|\frac{\partial \varphi}{\partial x}(x_{k})\right|,
\end{equation}
is always positive for the onion map if $h=0.97$ and $\gamma\in[0.15,1.85]$. If we change the value of $h$, then the range of $\gamma$ for which $\lambda>0$ is slightly different. A positive value of the Lyapunov exponent implies {\it chaotic} behavior of the map reflected by the {\it butterfly effect}, i.e. strong sensitivity to the initial conditions. LE measures rate of separation of two trajectories of a chaotic map and its value depends only on the parameter.\\
A function critical in the analysis of chaotic systems is the Sinai-Ruelle-Bowen (SRB) density $\rho$, which contains statistical information of dynamics described by $\varphi$ \cite{young-statistics, young-srb, crimmins-srb, sliwiak-1d}. Intuitively, $\rho$ can be viewed as the likelihood of the trajectory passing through a non-zero-volume region of the manifold and, if normalized, $\rho$ can be viewed as a probability density function.  In case of one-dimensional maps defined on $[0,1]$, the SRB density $\rho$ is a function that maps $[0,1]$ to the set of non-negative real numbers, which satisfies the unity axiom, i.e., $\int_{0}^{1}\rho(x)\,dx=1$. Figure \ref{fig:hybrid_density} shows $\rho$ generated for the onion map at different values of the exponent $\gamma$. 
We observe the density distribution is smooth if $\gamma \leq 0.4$. When the exponent $\gamma$ becomes higher, but is still no larger than 1, the density function is clearly bounded, but have some non-smooth regions. If $\gamma \geq 1$, the $\rho$ distribution features discontinuous regions.

\begin{figure}
    \centering
    \includegraphics[width=1.\textwidth]{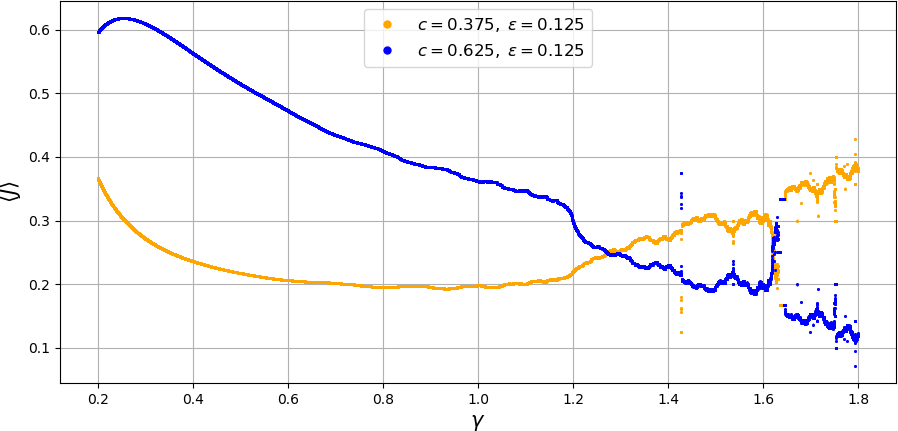}
    \caption{Relationship between the long-term average and the exponent $\gamma$ for the onion map (Eq. \ref{eqn:hybrid}) at $h=0.97$ with $J(x)=\delta_c^{\epsilon}(x)$. To generate this plot, we computed density distributions at a uniform grid of 16,001 different values of $\gamma$ between 0.2 and 1.8. For each value of $\gamma$, we run 10 indepentend simulations using $N = 41,943,040,000$ samples per simulation. In the calculation of the density, we divided the domain $x\in[0,1]$ into $K = 4$ bins of equal width (see Figure \ref{fig:hybrid_density} for reference).} 
    \label{fig:hybrid_statistics_panoramic}
\end{figure}

Due to the ergodicity of the onion map, its statistics does not depend on initial conditions. Moreover, this property implies that the Ergodic Theorem holds, and thus we can directly use the stationary density to compute the long-time averages of the onion map. In particular, the theorem ensures that an infinite time average of some quantity of interest $J$ is equal to the expected value of the same quantity computed with respect to the density distribution. Mathematically, it means that
\begin{equation}
    \label{eqn:ergodicity}
    \langle J\rangle = \lim_{N\to\infty}\frac{1}{N}\sum_{i=0}^{N-1}J(x_{i}) = \int_{0}^{1}J(x)\rho(x)dx
\end{equation}
always holds. It is assumed that $J$ is an integrable bounded function and it does not depend on the map parameter. The origins of the first assumption will be explained in Section \ref{sec:density-gradient}. Furthermore, the dependence of the objective function on the parameter is not considered in this paper, as it does not impose extra mathematical complexity in the computation of sensitivities and is irrelevant in the context of our analysis. Therefore, two critical properties can be further inferred from Eq. \ref{eqn:ergodicity}. First, the map statistics $\langle J\rangle$ solely depends on its parameter $\gamma$ and, second, the smoothness of statistics strictly depends on the smoothness of the density distribution. In our numerical test, we set $J(x)=\delta_c^\epsilon(x)$, $c\in[0,h]$, where $\delta_c^\epsilon(x)$ is an indicator function, i.e. $J(x)=1$ for all $x\in x_c^\epsilon:=[c-\epsilon/2,c+\epsilon/2]$, and $J(x)=0$ otherwise. 
Note with this particular choice of the quantity of interest, the long-time average equals the density distribution itself evaluated at $c$ in the limit $\epsilon\to 0$, i.e. $\langle J\rangle = \rho(c)$ if $\epsilon$ is infinitesimally small. 
For any $\epsilon$ such that $x_c^{\epsilon}\subset [0,1]$, $\langle J\rangle$ equals the integral of $\rho$ over $x_c^{\epsilon}$. Note also that for any Riemann-integrable $J(x)$, the statistics can be easily computed using a numerical integration scheme by virtue of Eq. \ref{eqn:ergodicity} if $\rho$ is available. Figure \ref{fig:hybrid_statistics_panoramic} illustrates the relationship between $\langle J\rangle$ and $\gamma$, for two different values of $c$. 

In terms of the function smoothness, we observe a similar trend in both Figure \ref{fig:hybrid_density} and Figure \ref{fig:hybrid_statistics_panoramic}. In particular, if $\gamma$ increases, both the density function and long-time average become more oscillatory and even discontinuous. This result is consistent with the study in \cite{blonigan-pdf}, where the relationship between the smoothness of the statistics and smoothness of the density distribution has been justified analytically using the Frobenius-Perron operator, which belongs to the class of Markov operators and describes the evolution of the SRB density \cite{ding-frobenius}. The authors of \cite{blonigan-pdf} notice that if the Frobenius-Perron operator is well-conditioned and the density function is differentiable in phase space (with respect to $x$ in 1D), then $\partial\rho/\partial\gamma$ must be bounded. This observation is critical for the existence of $d\langle J\rangle/d\gamma = \int_{0}^{1}J(x)\frac{\partial\rho}{\partial\gamma}(x)dx$, which is the sought-after quantity in sensitivity analysis.

Motivated by the above discussion, we strive to find a computable and generalizable mathematical criterion for the existence of $d\langle J\rangle/d\gamma$. In particular, our purpose is to identify a condition that can be translated to an efficient numerical method, and is applicable to higher-dimensional chaotic systems.

\section{Density gradient function in one-dimensional chaos}\label{sec:density-gradient}
\subsection{Density gradient function as an indicator of the differentiability of statistics of 1D Chaos}\label{sec:density-gradient-analysis}
The main focus of this section is to highlight the significance of the density gradient function $g$ in the context of the differentiability of statistics in one-dimensional chaotic maps. This function is a fundamental ingredient of the derivative of statistics. Here we consider 1D chaotic maps in which case $g$ is defined as follows, 
\begin{equation}
    \label{eqn:density-gradient-function}
        g(x) = \dfrac{d \log\rho}{dx}(x) = \frac{\rho'(x)}{\rho(x)}.
\end{equation}
That is, the density gradient $g$ is the relative rate of change of the SRB density at each point on the 1D manifold \cite{sliwiak-1d}. Throughout this section, we use the prime symbol ($'$) to indicate differentiation with respect to phase space. We assume $\varphi:[0,1]\to[0,1]$ is a 1D, invertible, ergodic, $C^3$ map with a positive LE. Let $J$ be a smooth observable whose expectation with respect to the SRB density or equivalently, the infinite-time average starting from almost everywhere, is denoted $\langle J\rangle.$ In this case, Ruelle's {\em linear response formula} \cite{ruelle-original, ruelle-corrections}, which is a closed-form expression for the parametric derivative of $\langle J\rangle$, is given by
\begin{equation} 
    \frac{d\langle J\rangle}{d\gamma}=\label{eqn:ruelle} 
    \frac{d}{d\gamma} \int_0^1 J(x) \;\rho(x)\,dx
    = \sum_{k=0}^{\infty} \int_0^1
    f(x)\,
    \big( J\circ\varphi_{k}\big)'(x) \;\rho(x)\,dx,
\end{equation}
where $f := \partial\varphi/\partial\gamma \circ \varphi^{-1}$ is the parameter perturbation. The subscript notation is used to denote the number of times a map $\varphi$ is applied i.e., $\varphi_0(x)=x$ and $\varphi_k(x) = \varphi(\varphi_{k-1}(x))$ for any state vector $x$, while the inverse of the map is indicated using the conventional notation, i.e.,  $\varphi^{-1}$. Integrating the RHS of Eq. \ref{eqn:ruelle} by parts leads to an alternative expression for the sensitivity, 
\begin{equation} 
        \label{eqn:s3}
        \frac{d}{d\gamma} \int_0^1 J(x) \;\rho(x)\,dx
        = -\sum_{k=0}^{\infty} \int_0^1 
        \Bigg(g(x)\;f(x)+f'(x)\Bigg) \, (J\circ\varphi_{k})(x)\;\rho(x)\,dx,
\end{equation}
which provides a direct relation between the derivative of the long-time average and the density gradient function $g$ (see \cite{sliwiak-1d} for the derivation of Eq. \ref{eqn:s3}). Eq. \ref{eqn:s3} is in fact a one-dimensional version of the space-split sensitivity (S3) formula, originally derived and computed in \cite{chandramoorthy-s3}. The general form of S3, which shows the relation $d\langle J\rangle/d\gamma$ vs. $g$ for higher-dimensional systems, is discussed in Section \ref{sec:lorenz-ruelle}.

The above formula is a sum of time-correlations between the function $J$ and $gf + f'.$ The $k$-time-correlation between two observables $\phi$ and $\psi$ is given by 
\begin{align}
\begin{split}
     C_{\phi,\psi}(k) = & \int_0^1 \phi\circ\varphi_k(x) \: \psi(x) \: \rho(x)\: dx - \\
    & \Big(\int_0^1 \phi(x)\: \rho(x) \: dx\Big) \Big(\int_0^1 \psi(x) \: \rho(x) \: dx \Big).
\end{split}
\end{align}
A classical result in uniform hyperbolicity theory is that $C_{\phi, \psi}(k)$ decays exponentially with $k>0$ for any pair of observables in a function class, at a uniform rate over that class, i.e., $C_{\phi,\psi}(k) \sim {\cal O}(e^{-ck})$ for some constant $c>0$ \cite{ruelle-original, chernov-decay}. The $k$-th term of regularized Ruelle's formula (Eq. \ref{eqn:s3}) is in fact a $k$-time correlation between $J$ and $h := gf + f'$ because 
\begin{align}
    \int_0^1 \Big(g(x)\:f(x) + f'(x)\Big) \: \rho(x) \: dx = \int_0^1 \Big(\rho(x)\:f(x)\Big)' \: dx = 0.   
\end{align}
The above integral vanishes since we have a periodic boundary. We remark that an analogous boundary term, which appears when we perform integration by parts on a higher-dimensional unstable manifold, also vanishes (see Section \ref{sec:lorenz-ruelle}). 
More generally, Ruelle's formula converges whenever $C_{J, h}(k)$ is summable. Our goal in this work is to investigate this condition so that we can assess the existence of linear response in systems that may not be uniformly hyperbolic. A numerical assessment is not only useful for practical purposes but necessary because mathematical analysis of such systems is difficult. We now isolate the term that determines the convergence. Considering again the $k$-th term of Ruelle's response,
\begin{align}
\begin{split}
\label{eqn:twoTerms}
    C_{J,h}(k) = & \int_0^1 
    J\circ\varphi_k  \: \Big(g(x)\:f(x) + f'(x)\Big)\:\rho(x)\: dx =\\ &
    \int_0^1 J\circ\varphi_k(x) \: g(x)\: f(x) \: \rho(x)\: dx  + \int_0^1 J\circ \varphi_k(x) \: f'(x)\:\rho(x) \: dx.
\end{split}
\end{align}    
In uniformly hyperbolic systems, the function class of observables exhibiting exponential decay of correlations contains $C^1$ functions. Hence, the time-correlation $C_{J, f'}(k)$, corresponding to the second integral in Eq. \ref{eqn:twoTerms}, is an exponentially decaying sequence in uniformly hyperbolic systems, since, by assumption $f \in C^2$. From here on, we discuss the choice of parameter perturbation, $f$, and the objective function, $J$,  such that $C_{J, f'}(k)$ is summable. Typically, in the context of engineering simulations, $J$ represents a physical quantity, such as force or temperature, which are smooth functions. The function $f$ and its derivative $f'$ are also smooth in case of many popular physical models, including the standard parameter perturbations in the Lorenz '63 system (see Section \ref{sec:lorenz}) \cite{lorenz-climate, sparrow-lorenz} or Kuramoto-Sivashinsky equation \cite{blonigan-ks}. 

We focus now on the first integral on the RHS of Eq. \ref{eqn:twoTerms}, isolate the components that are intrinsic to the dynamics, and whose convergence affects the existence of linear response. Since both $f g$ and $J$ are sufficiently regular, we expect that $C_{J, fg}(k)$ decays exponentially in uniformly hyperbolic systems. 
However, Ruelle's series may converge under weaker conditions on $g$. For instance, assuming that $C_{J,f'}(k)$ is absolutely summable, the boundedness of Ruelle's series depends only on the absolute summability of the correlation $C_{J, gf}(k)$. Now, to isolate conditions on $g$, we assume that $\langle J\rangle = 0.$ This assumption is without loss of generality because for any $J,$ $d_\gamma \langle J\rangle = d_\gamma \langle J - \langle J\rangle \rangle,$ and hence analyzing the existence of the derivative of $\langle J - \langle J\rangle \rangle$ is sufficient to determine the validity of linear response. Then, note that there exists some constant $c > 0,$ such that, for all $K \in \mathbb{Z}^+$, 
\begin{align}
\begin{split}
     &\left|\sum_{k\leq K} \Bigg( \int_0^1 J\circ\varphi_k(x) \: g(x)\: f(x) \:\rho(x)\:  dx + \int_0^1 J\circ\varphi_k(x) \: f'(x) \: \rho(x) \: dx\Bigg)\right|  \\ &\leq  
     \sum_{k\leq K}\big( |C_{J,gf}(k)|  + |C_{J,f'}(k)| \big) 
     \leq \sum_{k\leq K}  |C_{J,gf}(k)| + 
     c.
\end{split}
\end{align}
 From the above expression, we can see that when $C_{J,gf}(k)$ is absolutely summable, Ruelle's series is finite. Then, linear response holds. While this series converges exponentially in uniformly hyperbolic systems, and hence linear response holds, $|C_{J,gf}(k)|$ may be summable beyond  uniformly hyperbolic systems. Our goal is to find conditions on $g$ that can be verified numerically, and under which the time-correlation $C_{J,gf}(k)$ is summable, for any smooth perturbation $f$. Since $f$ can be considered to be arbitrarily smooth, a sufficient condition for summability is that $g$, and hence $f\: g$, must belong to a function class of observables for which exponential decay of correlations holds for any pair of observables belonging to this class. Exponential decay also indicates that a central limit theorem (CLT) of the following form also holds for the observables $h, h\circ\varphi, h\circ\varphi_2,...$, when $h$ is in the same function class \cite{young-statistics, liverani-decay, chernov-decay}. This CLT says that the random variable  $\frac{1}{\sqrt{N}}\sum_{n=0}^{N-1} \big( h\circ\varphi_n(x) - \langle h\rangle\big)$ is, for large $N$, and almost every $x$, distributed according to a normal distribution with mean 0 and a finite variance.

The crux of our assessment lies in our efficient numerical method to compute $g$ along a $\rho$-typical trajectory. With the values of $g$ available along a sufficiently long trajectory, we numerically assess whether the CLT applies by checking if the first and second moments of $|g|$ exist. We expect that when the CLT does apply to $|g|$, linear response holds. Accordingly, in the numerical examples in Sections \ref{sec:onion-results}-\ref{sec:lorenz}, we find that the validity of CLT for $|g|$ is indeed a sufficient condition. Moreover, we observe that a necessary and sufficient condition, in the examples considered, is that the first moment of $|g|$ is finite, i.e., $g \in L^1(\rho).$ Since an analytical assessment of the decay of correlations and CLT, beyond some hyperbolic systems \cite{young-statistics}, is absent, if this relationship between the existence of moments of $|g|$ and that of linear response is generalizable, we may be able to numerically determine the validity of linear response. Next we describe our numerical method to compute $g$ along trajectories.

\subsection{Computing the derivative of density $\rho'$ and density gradient $g$}\label{sec:algorithm}

\begin{figure}
    \centering
    \includegraphics[width=1.\textwidth]{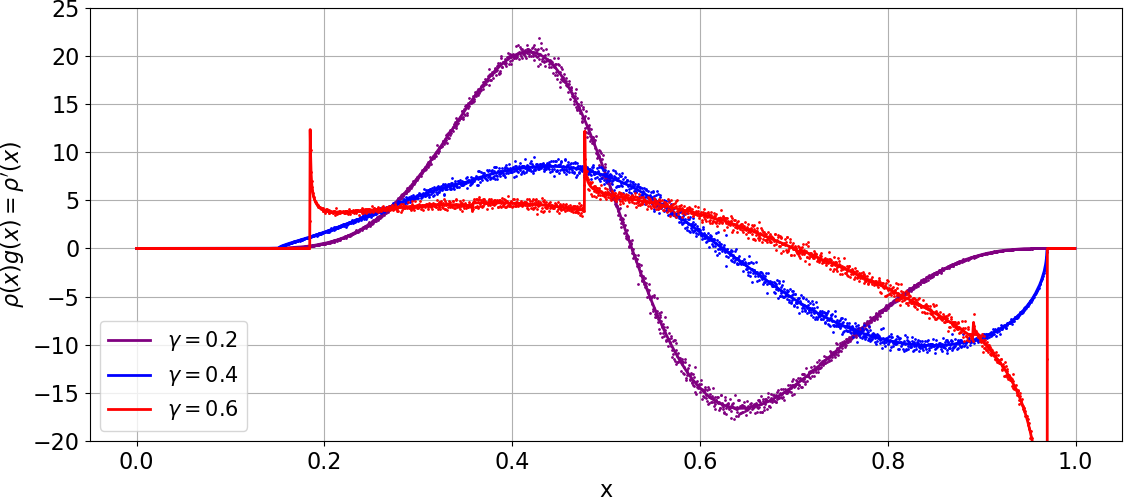}

    \centering
    \includegraphics[width=1.\textwidth]{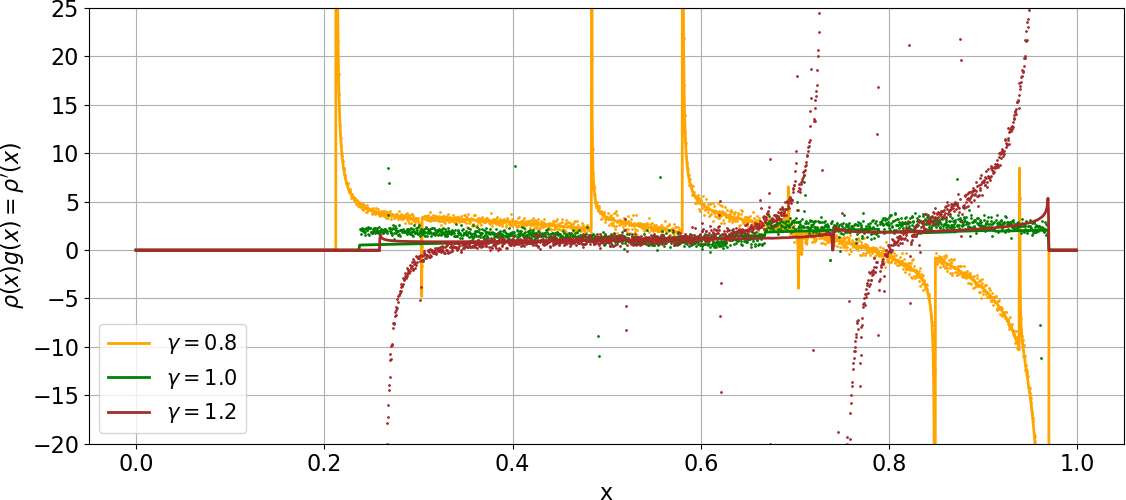}

    \caption{Derivative of density of the onion map (Eq. \ref{eqn:hybrid}) at $h=0.97$. We used $N=41,943,040,000$ samples and $K = 2048$ bins to generate all curves. The solid lines represent the derivative of density computed using Eq. \ref{eqn:density-gradient-iterative}-\ref{eqn:g-approximation}, while the dots represent central finite difference approximation of the same function using the corresponding density function histograms illustrated in Figure \ref{fig:hybrid_density}.}
    \label{fig:onion_g}
\end{figure}

In this section, we focus on computational aspects of the density gradient function. Due to the fact that the SRB density, $\rho$, is stationary in time, it satisfies 
\begin{equation}
    \label{eqn:frobenius-perron}
    \rho(\varphi(x)) = \frac{\rho(x)}{|\varphi'(x)|}.
\end{equation}
This statement is an alternative expression of measure preservation,
which implies that for any $J \in L^1(\rho),$
\begin{align}
\label{eqn:measure-preservation}
    \int_0^1 J(x) \: \rho(x)\: dx = \int_0^1 J\circ\varphi(x)\: \rho(x) \: dx.     
\end{align}
Applying a change of variables $x\to \varphi(x)$ to the integral on the right hand side, and recognizing that Eq. \ref{eqn:measure-preservation} holds for all $J \in L^1(\rho)$, leads to Eq. \ref{eqn:frobenius-perron}.  
Taking the logarithm of Eq. \ref{eqn:frobenius-perron} and then differentiating with respect to $x,$ we obtain,
\begin{equation}
        \label{eqn:density-gradient-iterative}
         g(\varphi(x)) = \frac{g(x)}{\varphi'(x)}-\frac{\varphi''(x)}{\varphi'(x)^{2}}.
\end{equation}
The above equation converges to the true value $\rho'(x)/\rho(x)$ upon iterating with an initial guess $g\circ\varphi_{-N}(x) = 0,$ as $N\to \infty$. A convenient way to numerically verify Eq. \ref{eqn:density-gradient-iterative} is to approximate $\rho' = \rho g$ using the above formula for $g$.  We validate the results against the finite difference approximation of $\rho'$. Let $x_0,x_1,...,$ be a long trajectory, 
and let the interval $[0,1]$ be divided into $K$ subintervals (bins), $\left\{\Delta_k\right\}_{k=1}^K,$ of equal length $1/K$.  We compute a piecewise-constant approximation of $\rho(x) g(x)$ as follows:
\begin{equation}
\label{eqn:g-approximation}
          \rho(x) g(x) \approx \dfrac{K}{N}\sum_{n=0}^{N-1} g(x_n) I_{\Delta_k}(x_n), \: \forall \: x \in \Delta_k
\end{equation}
where $I_A$ is the indicator function over a subset $A\subset [0,1]$. That is, $I_A(x) = 1,$ when $x \in A,$ and $I_A(x) = 0,$ otherwise. The pointwise error associated with this approximation is proportional to $\sqrt{K/N}$, i.e., $|\rho(x)g(x) - (K/N)\sum_{n=0}^{N-1} g(x_n) I_{\Delta_k}(x_n)|$ decays as $\mathcal{O}(\sqrt{K/N})$ for all $x\in \Delta_k$ if $g$ obeys the CLT \cite{chernov-decay}. Note that, for a fixed $N$, the error increases proportionally to $\sqrt{K}$ for all $x\in[0,1]$, because the approximation is piecewise-constant on a uniform grid of size $1/K$. Note if we replace $g(x_n)$ with 1 in the RHS of Eq. \ref{eqn:g-approximation}, we effectively obtain a formula for the density function itself. Thus, from the algorithmic point of view, the process of generating $\rho'$ requires similar steps as the process of generating $\rho$, while $g$ emerges as a byproduct. Analogously, this process can be generalized to higher-dimensional systems with a 1D unstable manifold. In such systems, $g$ is a scalar function, and thus Eq. \ref{eqn:g-approximation} still applies assuming an analogous partition of the higher-dimensional attractor is created. Computational aspects involving the density gradient function defined for higher-dimensional maps with one positive LE are described in Section \ref{sec:lorenz-ruelle} and \ref{app:density-gradient}.\\
Figure \ref{fig:onion_g} illustrates the derivative of density generated for the onion map for the same set of parameter values as the densities in Figure \ref{fig:hybrid_density}. We observe a satisfactory match between the results generated using the above algorithm for $\rho'(x)$ and the corresponding finite difference approximations as long as $\gamma<1.0$. For larger values of $\gamma$, there is a visible discrepancy between the two approximations in the proximity of discontinuities, which is consistent with the density $\rho$ exhibiting discontinuities for $\gamma > 1.0$ (compare with Figure \ref{fig:hybrid_density}).


\section{Probing the differentiability of statistics in one-dimensional chaos}\label{sec:onion-results}
\subsection{Analysis of the distribution of $|g|$}
The central part of this work is to identify a computable mathematical criterion for the differentiability of statistics. As argued in Section \ref{sec:density-gradient-analysis}, we anticipate there is a relation between the properties of the distribution of $|g|$ and the validity of linear response. We intend to further investigate this observation using numerical simulation. We compute the distribution of the absolute value of the density gradient (using Eq. \ref{eqn:density-gradient-iterative}) over a range of parameter values. We also compute the statistics-vs-parameter curve directly, from which we estimate its H\"older exponent. Very long trajectories are required to make an accurate estimation. However, it is still computationally feasible given the low dimensionality of our examples. Using this direct check of differentiability, we find any potential correlation between the probability distribution of $|g|$ and the existence of the parametric derivative.

\begin{figure}
    \centering
    \includegraphics[width=1.\textwidth]{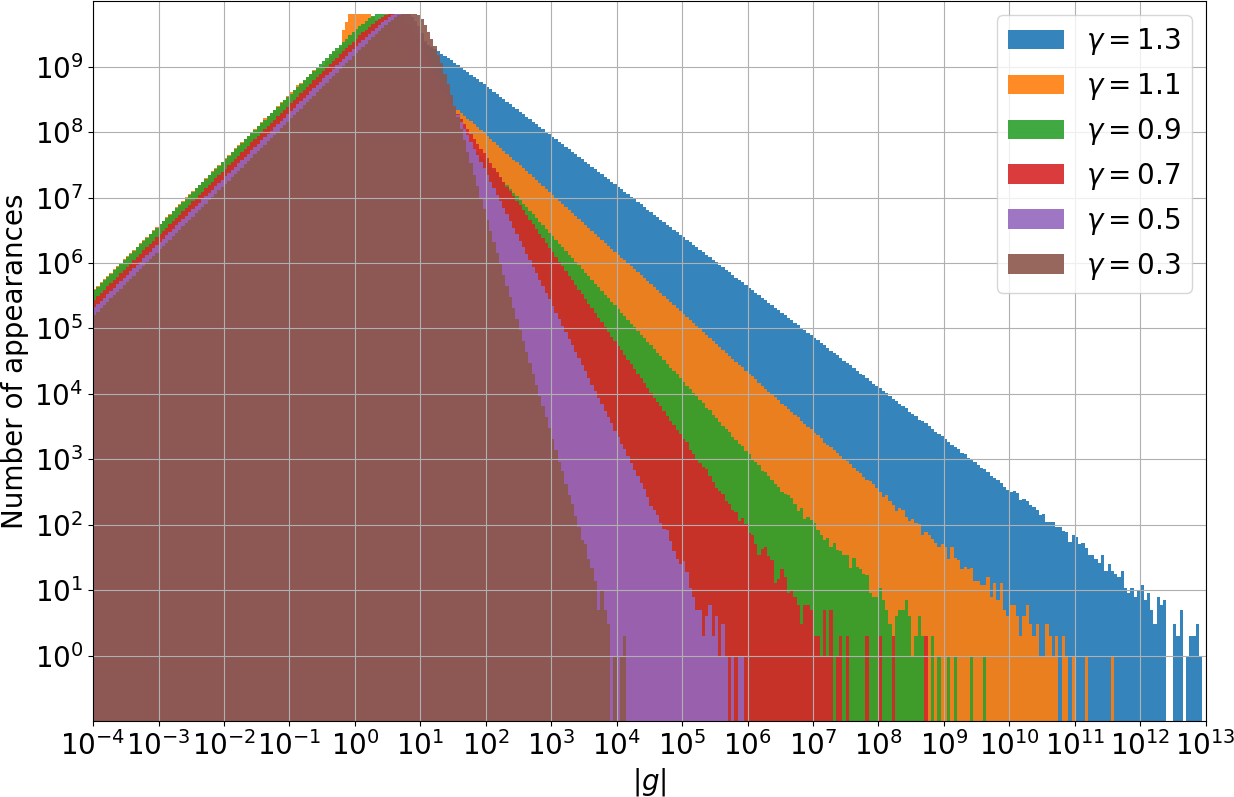}
       \caption{Distribution of the absolute value of the density gradient function generated for the onion map (Eq. \ref{eqn:hybrid}) at $h=0.97$. To generate these histograms, we divided the x-axis from $10^{-18}$ to $10^{84}$ into $K = 2048$ bins with equal with in the logarithmic scale. For each histogram, a trajectory of the length of approximately $N = 1.25\cdot10^{11}$ has been computed.}
    \label{fig:onion_tail}
\end{figure}

As the first step, we study the distribution of $|g|$, considering it to be a random variable, for the onion map introduced in Section \ref{sec:onion}. We obtain empirically the distributions of $|g|$ at different $\gamma$ values, from both the smooth and non-smooth regions. Based on the procedure introduced in Section \ref{sec:algorithm}, we compute a sufficiently long trajectory, and count the number of occurrences of $|g|$ in all bins, each corresponding to a subset of the range of $|g|$. Figure \ref{fig:onion_tail}  shows the distribution of $|g|$ on a logarithmic scale at fixed $h=0.97$ for different values of the parameter $\gamma$. 

The vertical axis of Figure \ref{fig:onion_tail} represents the number of appearances of a given value of $|g|$ in each bin. Based on these histograms, we conclude that the probability density function (PDF) of $|g|$, denoted as $\mathrm{PDF}(|g|)$, has power-law behavior, i.e. $\mathrm{PDF}(|g|)\sim |g|^{-t}$, for some exponent $t$. Since Figure \ref{fig:onion_tail} presents data on a log-log scale, the bin size increases proportionally to the value of $|g|$. Therefore, in Figure \ref{fig:onion_tail}, we observe a distribution that is proportional to $|g|\cdot\mathrm{PDF}(|g|)\sim|g|^{-t+1}$. That is, the slope of the histograms gives us the value of $-t +1$. 

For power-law probability distributions, the existence of the expected value, variance, and higher-order moments, is solely determined by the value of the exponent $t$. If $t\leq 2$, then the mean (expected value), $$\mathbb{E}[|g|]=\int_0^\infty |g|\: \mathrm{PDF}(|g|)\: d|g| = \int_0^1 |g(x)|\: \rho(x) \: dx,$$ 
and all higher moments are infinite. If $t>2$, then the mean is finite. In other words, the density gradient belongs to $L^1(\rho)$, when the slope of the histogram in Figure \ref{fig:onion_tail} is less than -1. In addition, if the exponent $t$ is larger than 3, then the variance $\mathrm{var}(|g|) = \mathbb{E}[g^2] - (\mathbb{E}[|g|])^2$ of the probability distribution function is finite. In this case, the density gradient is square-integrable with respect to $\rho$, i.e., $g\in L^2(\rho)$.

From Figure \ref{fig:onion_tail}, the slope at $\gamma \leq 0.9$ is always less than -1. Thus, the absolute value of the density gradient has finite expected value, and $g$ is therefore Lebesgue-integrable as long as $\gamma \leq 0.9$. Thus, the Lebesgue-integrability threshold must be in $(0.9, 1.1)$, as all the distributions corresponding to $\gamma \geq 1.1$ have slopes $t$ larger than $-1$. Furthermore, square-integrabilty threshold can be estimated to be around $\gamma = 0.5$, since the slope at $\gamma = 0.5$ is approximately equal to -2. For smaller values of $\gamma$, for example, $\gamma = 0.3$, we see that both expectation and variance are finite.

\subsection{H\"older exponent test}\label{sec:holder}
In order to draw a correlation, if any, between the existence of moments of $|g|$ and the validity of linear response, we must identify the regions of parameter space where linear response exists.
For this, we directly assess the smoothness of the statistics illustrated in Figure \ref{fig:hybrid_statistics_panoramic} for the onion map. That is, we numerically estimate the H\"older exponent $\mu\in(0,1]$ of the long-time average function, which changes with $\gamma$. 
A function $h(\gamma)$ defined on a domain $D$ in parameter space is said to be H\"older continuous with exponent $\gamma$, if there exists a $C > 0$ such that
\begin{equation}
\label{eqn:holder-inequality}
|h(\gamma_1)-h(\gamma_2)|\leq C|\gamma_1-\gamma_2|^{\mu}
\end{equation}
for all $\gamma_1$ and $\gamma_2$ in $D$
If $\mu = 1$, then $h(\gamma)$ is Lipschitz-continuous, and in this case, also differentiable at almost every parameter value in $D.$ 
Thus, to probe the smoothness of the long-time average, we numerically estimate the H\"older exponent of the statistics-parameter relation, $\langle J\rangle$ vs. $\gamma$, from the plot in Figure \ref{fig:hybrid_statistics_panoramic}.
This can be achieved by generating a sufficient number of data points and producing a scatter plot with $|\langle J\rangle(\gamma_1)-\langle J\rangle(\gamma_2)|$ on the $y$-axis and $|\gamma_1-\gamma_2|$ on the $x$-axis, where $\gamma_1$ and $\gamma_2$ indicate points of evaluation of $\langle J\rangle(\gamma)$. If the logarithmic scaling is used, the H\"older exponent $\mu$ can be approximated by estimating the slope (steepness) of the maximum values of $|\langle J\rangle(\gamma_1)-\langle J\rangle(\gamma_2)|$ as $|\gamma_1-\gamma_2|$ changes. 

Assuming the function $\langle J\rangle$ is sampled every $\delta\gamma$ along the x-axis, it is clear that $\delta\gamma = \min_{\gamma_1,\gamma_2\in D}|\gamma_1-\gamma_2|$. We set $\delta\gamma=0.0001$, which allows us to capture high-frequency oscillations. The value of $\delta\gamma$, however, cannot be too small, as the growing statistical noise may significantly impact the value of $|\langle J\rangle(\gamma_1)-\langle J\rangle(\gamma_2)|$. To further reduce the effect of statistical noise, we run 10 independent simulations per one parameter value and compute the 3-sigma confidence interval of the data coming from these independent simulations, where the standard deviation is averaged over a chosen interval of $\gamma$. 

\begin{figure}
    \centering
    \includegraphics[width=0.48\textwidth]{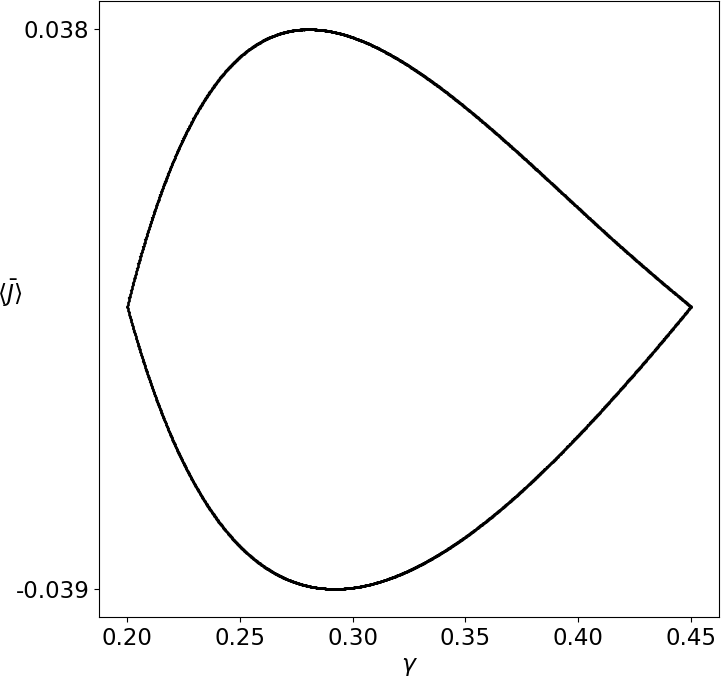}
    \includegraphics[width=0.48\textwidth]{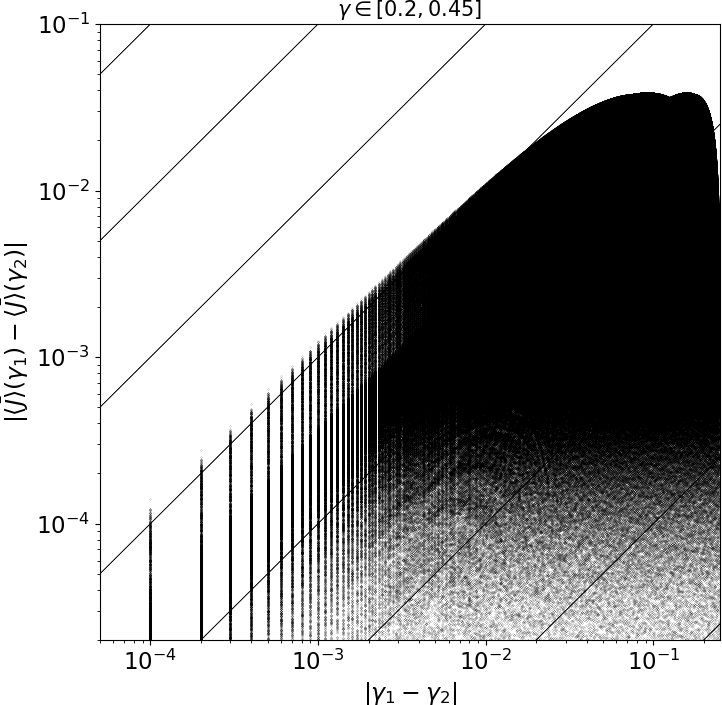}
    \\
    \vspace{5mm}
    \includegraphics[width=0.48\textwidth]{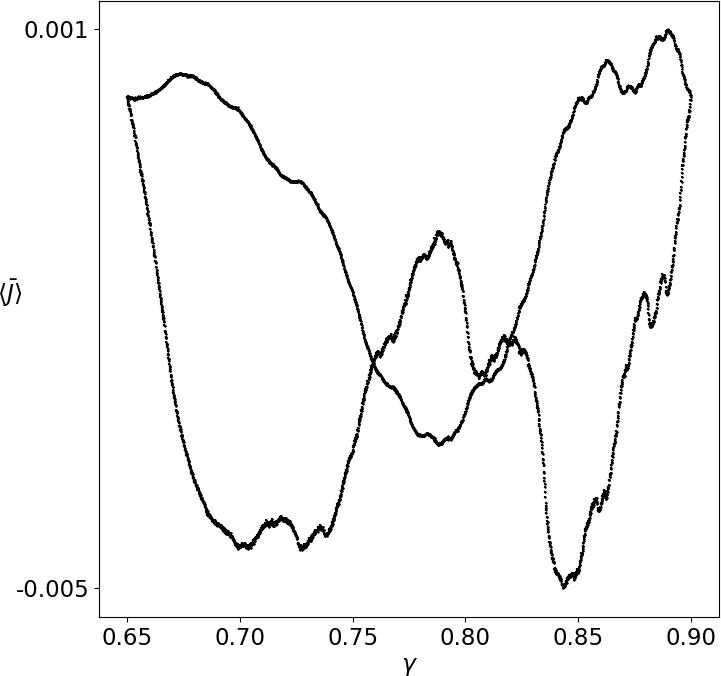}
    \includegraphics[width=0.48\textwidth]{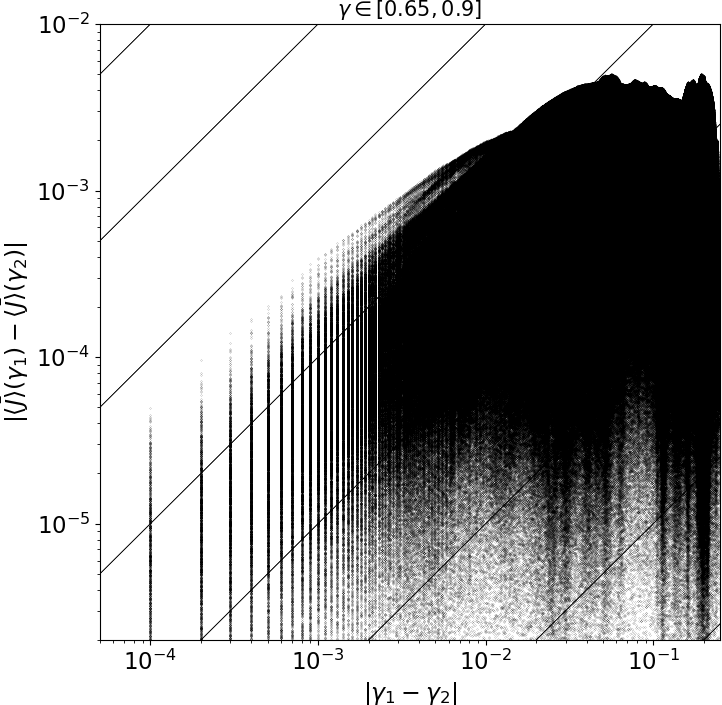}
    \caption{Left column: relation of the long-time average and the exponent $\gamma$ for the onion map at $h=0.97$. The simulation data is the same as the data presented in Figure \ref{fig:hybrid_statistics_panoramic} for $c = 0.625$, however the quantity of interest has been changed to $\langle \bar{J}\rangle$ by subtracting a linear function describing a straight line crossing the endpoints of the curve in Figure \ref{fig:hybrid_statistics_panoramic} in each $\gamma$ interval from the original objective function. Each plot corresponds to a different $\gamma$ interval between $\gamma_{min}$ and $\gamma_{max}$, which has been discretized uniformly with step size $\delta\gamma=0.0001$. For each value of $\gamma$, we run 10 independent simulations. Right column: H\"older exponent test results. First, for each pair of data points from the left-hand side plot, excluding the pairs with the same value of $\gamma$ (i.e. when $\gamma_1=\gamma_2$), we compute the difference of the corresponding long-time average values versus the difference of their parameter values. Second, we compute the lower-bound of the 3-sigma confidence interval by subtracting 6 averaged sigmas, where sigma represents standard deviation of results obtained in 10 simulations averaged over the interval $[\gamma_{min},\gamma_{max}]$, from the computed differences of modified statistics. This means each plot has approximately $(0.5\cdot(\gamma_{max}-\gamma_{min})/\delta\gamma)^2\approx 1.6\cdot 10^6$ data points. Skew solid lines represent reference lines with the slope of 1 in the logarithmic scale.}
    \label{fig:onion_holder1}
\end{figure}

\begin{figure}
    \centering
    \includegraphics[width=0.48\textwidth]{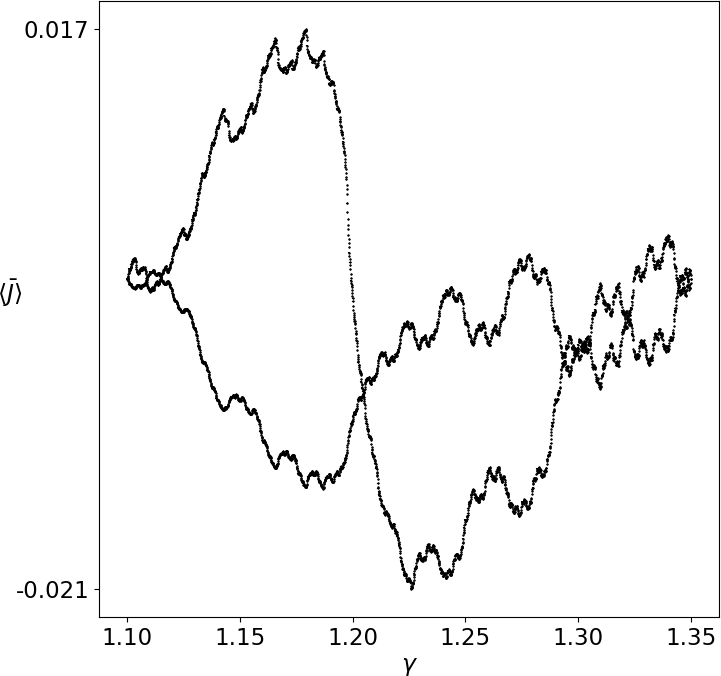}
    \includegraphics[width=0.48\textwidth]{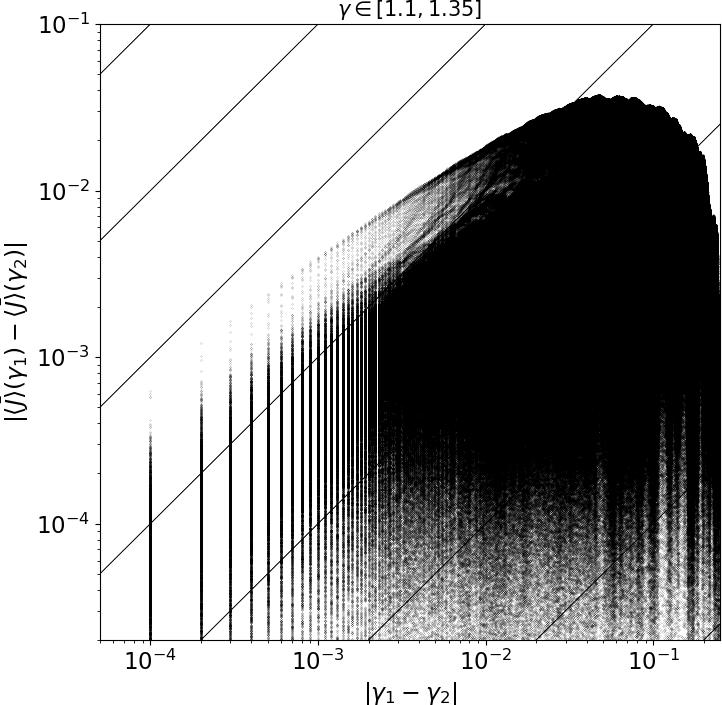}
    \\
    \vspace{5mm}
    \includegraphics[width=0.48\textwidth]{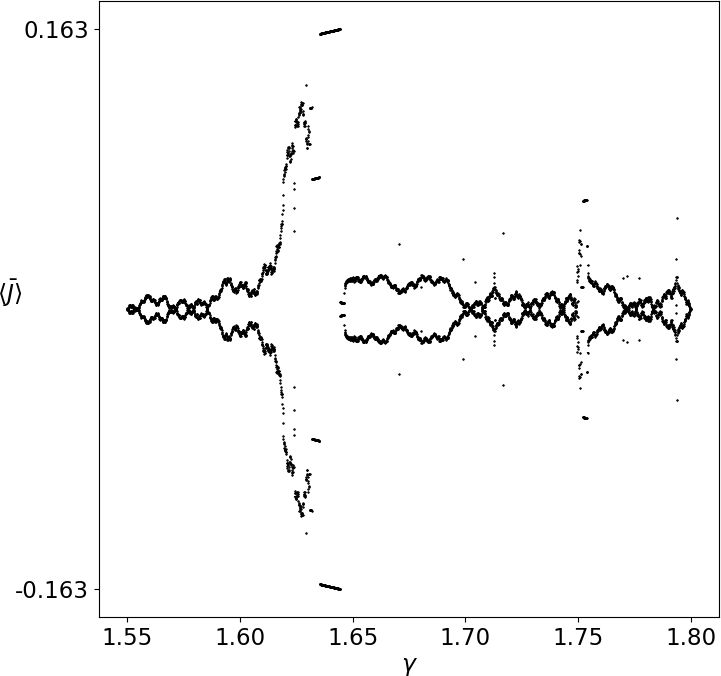}
    \includegraphics[width=0.48\textwidth]{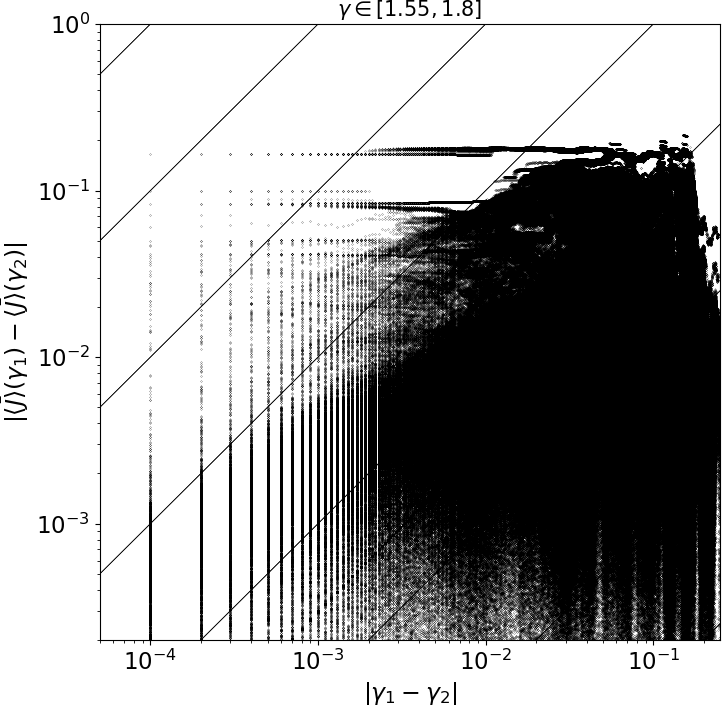}
    \caption{This figure is an extension of Figure \ref{fig:onion_holder1}. It includes $\gamma$ intervals corresponding to non-smooth statistics. All plots have been generated in the same manner as those in Figure \ref{fig:onion_holder1} -- see caption therein for more details.}
    \label{fig:onion_holder2}
\end{figure}

The left-hand side column of Figures \ref{fig:onion_holder1}  and \ref{fig:onion_holder2} illustrate the statistics versus parameter dependence at four different intervals of $\gamma$. The second column of these two figures shows $|\langle \bar{J}\rangle(\gamma_1)-\langle \bar{J}\rangle(\gamma_2)|$ versus $|\gamma_1-\gamma_2|$ computed from the data presented in the left-hand side column, where $\langle\bar{J}\rangle$ represents the long-time average of a modified objective function $\bar{J}$. The new quantity of interest is obtained by subtracting a linear function from $\langle J\rangle$, illustrated in Figure \ref{fig:hybrid_statistics_panoramic}, such that the resulting long-time average vanishes at the end points of each interval of $\gamma$. This modification is made to visually amplify the roughness of the curve, which is done for demonstration purposes only. See the caption of Figure \ref{fig:onion_holder1} for more details.

The top row of Figure \ref{fig:onion_holder1} corresponds to the range $\gamma\in[0.2,0.45]$. It is evident that the long-time average is smooth in this interval, as it satisfies Ineq. \ref{eqn:holder-inequality} with the exponent $\mu\approx 1$. According to Figure \ref{fig:onion_tail}, the tail of the distribution of $|g|$ has a slope smaller than -2 in that interval, which implies that $|g|$ has both finite mean and variance. The second row of Figure \ref{fig:onion_holder1} corresponds to the interval $\gamma\in[0.65,0.9]$, in which the statistics seems sharper, but the H\"older exponent $\mu$ is still close to 1. Figure \ref{fig:onion_tail} indicates that $g$ is in $L^1(\rho)$ but not in $L^{2}(\rho)$ in that range (it has finite mean, but infinite variance), as the slope of the distribution tail is between -2 and -1. Figure \ref{fig:onion_holder2} includes two sets of plots showing clearly non-smooth, even discontinuous responses. Even for the interval $\gamma\in[1.1,1.35]$, the H\"older exponent is significantly smaller than 1, which indicates that the statistics are not differentiable. In case of $\gamma\in[1.55,1.8]$, the long-time average is not even H\"older-continuous with respect to $\gamma$. Again correlating with our numerical results on the distribution of $|g|,$ for $\gamma > 1$, we found that $|g|$ does not have   a finite mean.
\noindent

The plots in the right column of Figures \ref{fig:onion_holder1}  and \ref{fig:onion_holder2} clearly indicate that the statistics of the onion map is differentiable as long as $\gamma$ is smaller than 1. 
According to our analysis of the distribution of $|g|$, $\gamma < 1.0$ implies finite expectation of $|g|$. It means that, in case of the onion map, linear response holds when $g\in L^1(\rho)$. This result also confirms our analysis from Section \ref{sec:density-gradient-analysis}. From our numerical results, in this case, we find that the converse is also true: when $g \notin L^1(\rho)$, linear response fails. To check whether the equivalence 
$$g\in L^1(\rho)\Longleftrightarrow \left|\frac{d\langle J\rangle}{d\gamma}\right|<\infty$$
is generalizable, we will apply the above two-step procedure to a higher-dimensional system with one positive LE.

\section{Generalization to a higher-dimensional system}\label{sec:lorenz}
\subsection{Lorenz '63 system}\label{sec:lorenz-presentation}
\begin{figure}
\begin{minipage}{0.57\textwidth}
    \centering
    \includegraphics[width=1.\textwidth]{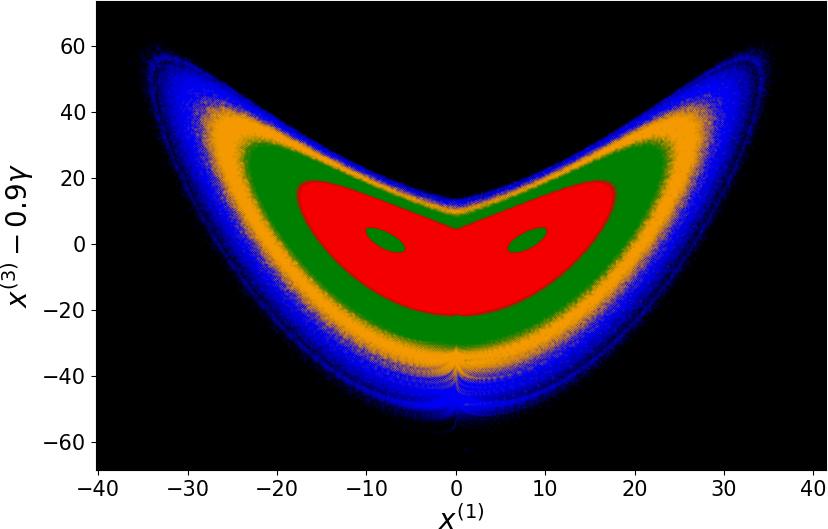}
\end{minipage}
\begin{minipage}{0.42\textwidth}
    \centering
    \includegraphics[width=1.\textwidth]{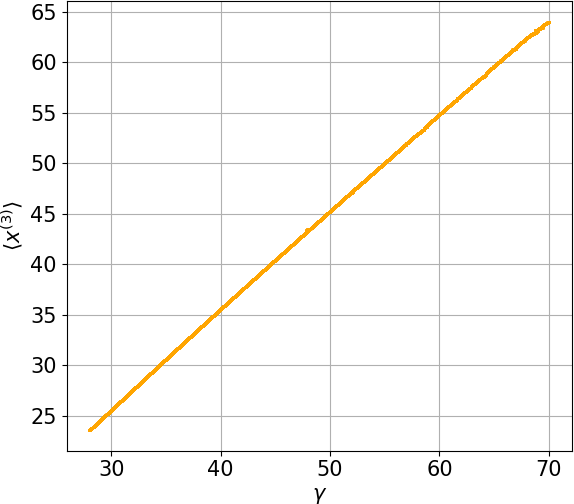}
\end{minipage}
    \caption{Left: projection on the $x^{(1)}$-$x^{(3)}$ plane of the Lorenz attractor at $\gamma =$ 25 (red), 40 (green), 55 (orange) and 70 (blue). Each projection has been shifted downwards proportionally to $\gamma$ for demonstration purposes. Right: relation of the long-time average $\langle x^{(3)}\rangle$ (defined by Eq. \ref{eqn:z_jav}) and the system parameter $\gamma$. We generate 420,000 data points in total: for each value of $\gamma$ on a uniform grid with size $\delta\gamma = 0.001$, we run 10 independent simulations. For each data point, we compute approximately $2.5\cdot10^{10}$ time steps with $\Delta t = 0.01$.}
\label{fig:lorenz_68}
\end{figure}

\begin{figure}
    \centering
    \includegraphics[width=1.0\textwidth]{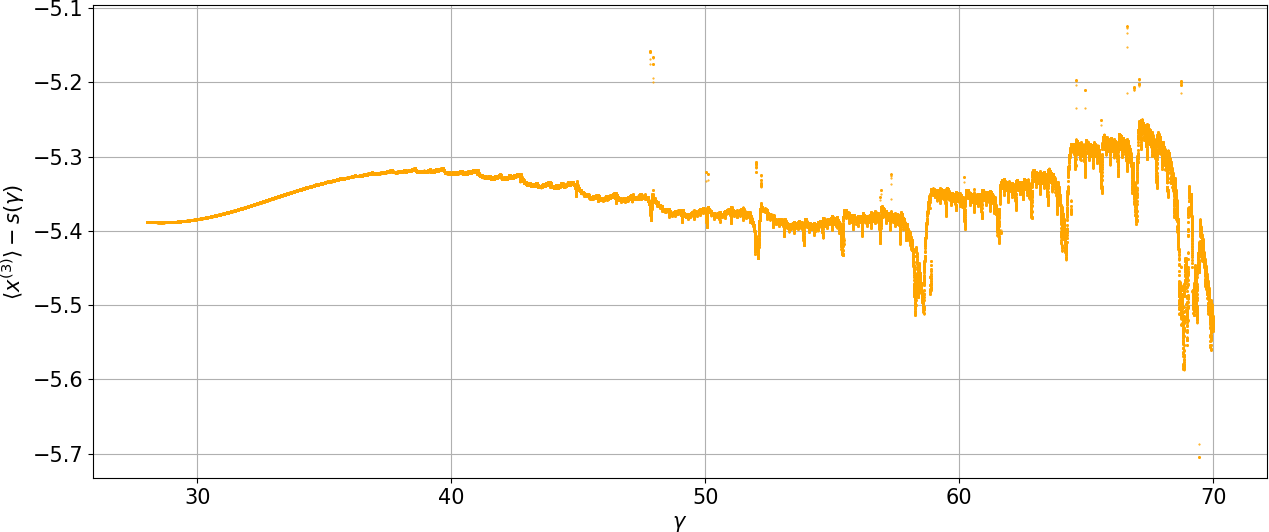}
    \caption{The modified quantity of interest, $\langle x^{(3)} \rangle - s(\gamma)$, as a function of $\gamma$, where $s(\gamma) = 1.06 \gamma - 0.00095 \gamma^2$. The quadratic function $s(\gamma)$ is a result of least squares polynomial fitting of the original long-time average shown in Figure \ref{fig:lorenz_68}.}
    \label{fig:lorenz_68_statistics}
\end{figure}

In this section, we generalize our conclusions from Section \ref{sec:onion-results} to higher-dimensional systems with a one-dimensional unstable manifold. This means we consider $n$-dimensional systems that have exactly one positive Lyapunov exponent out of $n$ Lyapunov exponents. As a test case, we consider the Lorenz '63 system \cite{lorenz-climate, sparrow-lorenz}, which consists of three coupled nonlinear ODEs,
\begin{equation}
\label{eqn:lorenz}
    \frac{dx^{(1)}}{dt} = \sigma(x^{(2)}-x^{(1)}),\;\; \frac{dx^{(2)}}{dt} = x^{(1)}(\gamma - x^{(3)}) - x^{(2)}, \;\; \frac{dx^{(3)}}{dt} = x^{(1)} x^{(2)} - \beta x^{(3)},
\end{equation}
where $\sigma \geq 0$, $\beta \geq 0$, and $\gamma \geq 0$ are the system parameters. The solution to Eq. \ref{eqn:lorenz} is represented by a 3-element state vector $x(t)=[x^{(1)}(t),x^{(2)}(t),x^{(3)}(t)]^T$. In our analysis, we set $\sigma$ and $\beta$ to their canonical values of 10 and 8/3, respectively, and keep them fixed, while we allow $\gamma$ to vary. Given the Lorenz '63 system is a three-dimensional system, it has three distinct Lyapunov exponents $\lambda_i$, $i= 1, 2, 3$, indexed in decreasing order. They satisfy the following constraints \cite{sparrow-lorenz},
\begin{equation}
\label{eqn:lorenz_le}
    \lambda_1 + \lambda_2 + \lambda_3 = -(1+\sigma+\beta),\;\;\;\;\lambda_2 = 0.
\end{equation}
Since both parameters are assumed to be positive, it is evident that Eq. \ref{eqn:lorenz_le} admits at most one positive solution. According to \cite{sparrow-lorenz}, for the canonical values of $\sigma$ and $\beta$, the Lorenz system is:
\begin{itemize}
    \item non-chaotic (has no positive LEs) if $0 \leq \gamma < 24.7$ and $\gamma > 99.5$,
    \item chaotic (has one positive LE) if $24.7\leq\gamma\leq 99.5$. 
\end{itemize}
Therefore, in this section, we focus on the smoothness of statistics of the Lorenz '63 system when $\gamma\in[24.7,99.5]$. To generate all results presented in this section, we integrate the system forward in time using the second-order Runge-Kutta scheme (midpoint method) with time step\footnote{Specific values of the time step size $\Delta t$ are indicated in the captions of corresponding figures.} $\Delta t$, starting from a random initial vector $x_{\mathrm{init}}$. In our discussion, we no longer consider the original, that is, continuous version of Lorenz '63, but rather we focus on the discrete form using a map $\varphi$, which is defined by the numerical time integration of the Lorenz '63 system for a time of $\Delta t$; that is, $\varphi(x(t)) = x(t + \Delta t),$ for all $t \in \mathbb{R}^+$. In other words, an orbit of $\varphi$, denoted $x_k,$ $k\in \mathbb{Z}^+$, is a numerical solution of the Lorenz '63 system with $x_k$ being the 3-element state vector at time step $k$. \\
As a quantity of interest, we consider the long-time average of the third variable,
\begin{equation}
    \label{eqn:z_jav}
    \langle x^{(3)} \rangle = \lim_{N\to\infty}\frac{1}{N}\sum_{k=0}^{N-1} x^{(3)}_k,
\end{equation}
which we approximate as $\langle x^{(3)}\rangle$ by generating sufficiently long trajectories. Figure \ref{fig:lorenz_68} shows a 2D projection of the attractor at different values of $\gamma$, as well as the dependence of $\langle x^{(3)}\rangle$ on $\gamma$. We observe that the attractor expands outward on the $x^{(1)}$-$x^{(3)}$ plane, as $\gamma$ increases. This observation is also reflected in the linear relation between $\langle x^{(3)}\rangle$ and $\gamma$, which is shown on the right-hand side of Figure \ref{fig:lorenz_68}. 
To show the statistical quantities of the Lorenz '63 system are in fact non-smooth at some values of $\gamma$, we subtract a smooth function $s(\gamma)$, obtained by fitting $x^{(3)}$ vs. $\gamma$ with a quadratic polynomial, from the original data shown in the right plot of Figure \ref{fig:lorenz_68}. Figure \ref{fig:lorenz_68_statistics} illustrates the behavior of the modified quantity of interest, i.e., $x^{(3)}-s(\gamma)$, as $\gamma$ changes. 
This computational treatment clearly reveals the actual regularity of the system's statistics. Analogously to the one-dimensional onion map, here as well we observe a transition from a smooth response at lower values of $\gamma$ to a non-smooth, and even discontinuous, behavior at larger values of $\gamma$. Note the modified statistics becomes sharp for values of $\gamma$ slightly above 30. This is in fact the region where the Lorenz system loses its quasi-hyperbolic properties \cite{sparrow-lorenz}.  

\begin{figure}
    \centering
    \includegraphics[width=0.7\textwidth]{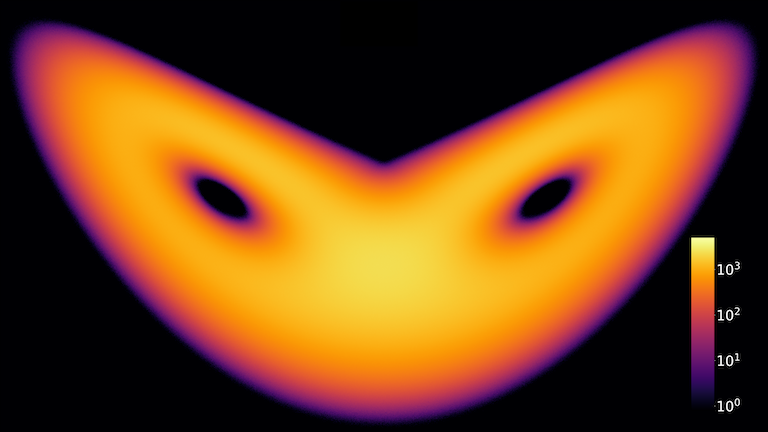}
    
    \includegraphics[width=0.7\textwidth]{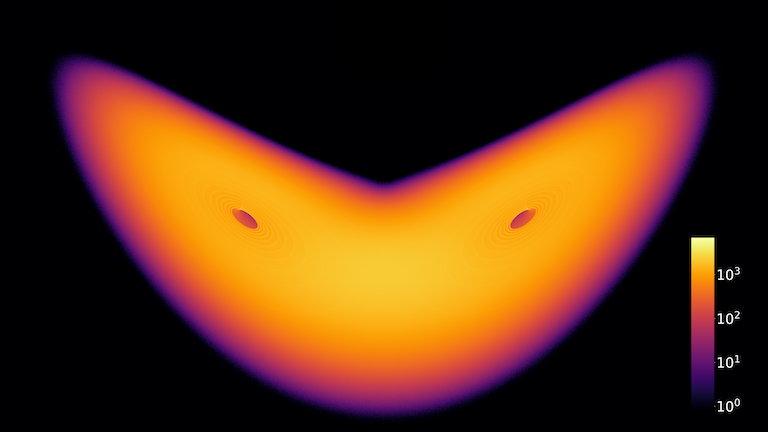}
    
    \includegraphics[width=0.7\textwidth]{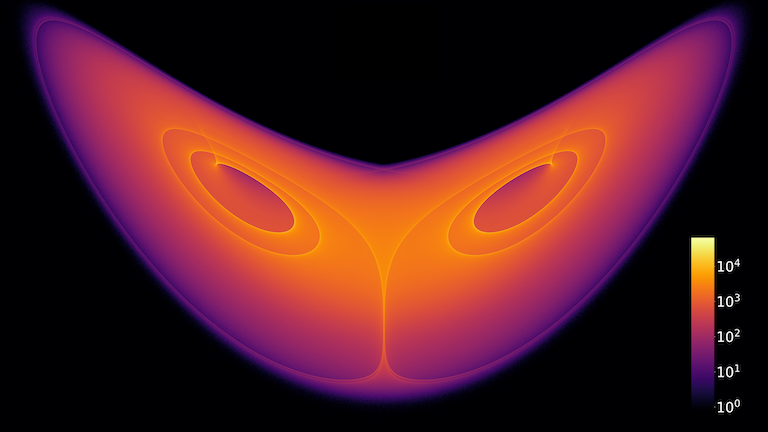}
    \caption{$x^{(1)}-x^{(3)}$ projection of the (unnormalized) density function of the Lorenz '63 system at $\gamma = 28$ (top), 38 (middle) and 70 (bottom). To generate each plot, a 2D box in phase space has been divided into $n_{x^{1}}\cdot n_{x^{3}}=3840\cdot 2160$ uniform rectangular cells/bins. The coordinates of the bottom left/upper right corner of each box are the following: $[-20,0]/[20,50]$ (top), $[-27,0]/[54,70]$ (middle), $[-40,0]/[40,80]$ (bottom). We computed a trajectory of length $N = 0.5\cdot 10^9$ with $\Delta t = 0.002$ for each plot. The color bars indicate the number of times the trajectory crosses a bin.}
    \label{fig:lorenz_68_density}
\end{figure}

For completeness, we illustrate the $x^{(1)}$-$x^{(3)}$ projection of the density function of the Lorenz system at three different values of $\gamma$ in Figure \ref{fig:lorenz_68_density}. We notice a clearly smooth distribution for $\gamma = 28$. For $\gamma = 38$, however, subtle wrinkles are visible around the ``eyes'' of the attractor. In case of $\gamma = 70$, regions with large density gradients, which clearly indicate non-smoothness of the distribution, appear around the ``eyes" and close to the boundary of the attractor. 

Based on these observations, we anticipate the density gradient function to be smooth for values of $\gamma$ close to 28, and non-smooth if $\gamma$ is higher. We also acknowledge a consistency between Figures \ref{fig:lorenz_68_statistics}-\ref{fig:lorenz_68_density} and Figures \ref{fig:hybrid_density}-\ref{fig:hybrid_statistics_panoramic}, corresponding to the Lorenz '63 system and onion map, respectively. Both pairs of figures indicate a strong correlation between the smoothness of statistics and smoothness of the density function in phase space. For a thorough investigation of this connection, we must numerically compute $g$, and also the H\"older exponents of the statistics-parameter response curve. The definition of $g$ in higher-dimensional systems and its impact on the sensitivity is discussed in the following section.

\subsection{Ruelle's formula, S3 and density gradient function in systems with one positive LE}\label{sec:lorenz-ruelle}
In this section, we consider Ruelle's formula \cite{ruelle-original,ruelle-corrections} as applied to multi-dimensional systems with one-dimensional unstable manifolds or equivalently, chaotic systems with one positive LE. We split Ruelle's formula into stable and unstable contributions out of which the unstable contribution is analogous to Eq. \ref{eqn:s3} for one-dimensional chaotic maps. Due to this similarity, we expect the relationship between the boundedness of linear response and the integrability of $g$ to also hold in multi-dimensional systems with a single direction of instability.

As before, we consider an invertible, ergodic, discrete map of a manifold $M$, parameterized by $\gamma$, and given by
\begin{equation}
   \label{eqn:e-system}
    x_{k+1} = \varphi(x_k;\gamma), \;\; k \in \mathbb{Z}.
\end{equation}
Here, $x_k$ is an $n$-dimensional state vector. Equation  \ref{eqn:e-system} may also arise from time-discretizations of continuous ODEs, such as the Lorenz system in Eq. \ref{eqn:lorenz}. Let $D$ denote the derivative operator with respect to phase space, and $D\varphi$ the $n\times n$ Jacobian matrix of the system. For the system defined by Eq. \ref{eqn:e-system}, Ruelle's formula \cite{ruelle-original, ruelle-corrections} for the parametric derivative of the long-time average can be expressed as 
\begin{equation}
    \label{eqn:e-ruelle}
    \frac{d\langle J\rangle}{d\gamma} = \sum_{k=0}^{\infty}\langle D(J\circ\varphi_k)\cdot\chi,\rho\rangle,
\end{equation}
where $\chi:=\partial\varphi/\partial\gamma \circ \varphi^{-1}$ is the parameter perturbation vector. This formula is proven \cite{ruelle-original} rigorously under the assumption of {\em uniform hyperbolicity} which guarantees the existence of uniformly expanding and contracting directions of perturbations. More precisely, in uniformly hyperbolic systems, at each $x \in M,$ there exists a decomposition of the tangent space, $T_x M = E^u(x) \oplus E^s(x),$ which satisfies the following properties: 
\begin{itemize}
    \item covariance property:
    $$ D\varphi (E^u(x)) = E^u(\varphi(x)),\; D\varphi (E^s(x)) = E^s(\varphi(x)),$$
    \item uniform expansion/contraction:\\
    for some fixed constants $C>0$, $\lambda\in(0,1)$, at every $x \in M$, every vector $v\in E^u(x)$ satisfies
    $$ \|D\varphi_{-k}(x)\;v(x)\| \leq C\lambda^k\|v(x)\|$$
    for all positive integers $k$. And, $$ \|D\varphi_{k}(x)\;v(x)\| \leq C\lambda^k\|v(x)\|$$ for all $v \in E^s(x).$ The norm, $\|\cdot \|$ denotes the standard Euclidean norm in $\mathbb{R}^n$. 
\end{itemize}
Although the series in Eq. \ref{eqn:e-ruelle} has been proven to converge, it is practically infeasible to compute it directly using tangent/adjoint methods \cite{lea-climate} in high dimensional systems. This computational infeasibility \cite{eyink-ensemble, chandramoorthy-turbulence} stems from the fact that the integrand $D(J \circ \varphi_k)\cdot \chi$ increases exponentially with $k$ for almost every perturbation $\chi$. In uniformly hyperbolic systems, it is possible to decompose the vector $\chi$ as $\chi_1 + \chi_2$ such that 
\begin{itemize}
    \item $\chi_1(x) \in E^u(x)$ and both components, $\chi_1$ and $\chi_2,$ are differentiable on the unstable manifold;
    \item there exists a bounded vector field $v:M \to \mathbb{R}^d$ that is orthogonal to $E^u$ and satisfies $v - D\varphi \: v = \chi_2.$
\end{itemize}
Given such a decomposition of $\chi$, we refer to the part of the sensitivity due to $\chi_2$ as the {\em stable} contribution, which is given by $\sum_{n=0}^\infty \langle DJ\cdot (D\varphi_{n} \chi_2) \circ \varphi_{-n}, \rho\rangle.$ It can be shown that the stable contribution is equivalently expressed as $\langle DJ \cdot v, \rho\rangle,$ including that $v := \sum_{n=0}^\infty \big(D\varphi_{n} \chi_2\big)\circ\varphi_{-n}$ is a bounded vector field \cite{chandramoorthy-s3}. Since, by assumption, $\|DJ\|_\infty$ is bounded, the stable contribution is always bounded.\\

We now turn our attention to the sensitivity due to $\chi_1,$ which we shall refer to as the {\em unstable} contribution: $\sum_{n=0}^\infty \langle D(J\circ\varphi_n) \cdot \chi_1, \rho\rangle.$ We shall restrict ourselves to one-dimensional unstable manifolds. Let $q(x)$ be the unit vector along the one-dimensional vector space, $E^u(x)$. Let $\xi$ be a coordinate along the unstable manifold, and let $\chi_1 = a \: q,$ where $a$ is the scalar field representing the component of $\chi_1$ along $q.$ We shall regularize the unstable contribution by applying integration by parts on the unstable manifold \cite{ruelle-corrections, jiang-srb} to yield,
\begin{align}
\label{eqn:unstableContribution}
    \sum_{n=0}^\infty \langle D(J\circ\varphi_n) \cdot \chi_1, \rho\rangle = - \sum_{n=0}^\infty \langle 
    (J\circ\varphi_n)\: (a g + b), \rho \rangle, 
\end{align}
where 
\begin{itemize}
    \item $b = \dfrac{\partial a}{\partial \xi},$ is the derivative of the vector field $\chi_1$ on the unstable manifold,
    \item $\rho^u$ is the density of the conditional measures of the SRB measure, $\rho,$ on unstable manifolds,
    \item $g := \dfrac{1}{\rho^u} 
    \dfrac{\partial \rho^u }{\partial \xi}$ is the density gradient function, equal to the derivative of the logarithm of $\rho^u$ on the unstable manifold.
\end{itemize}
The sum of the stable and unstable contributions, as defined above, specializes the S3 formula to systems with one positive LE. In the case $M$ itself is a one-dimensional manifold, as in the onion map, the unstable contribution, given by Eq. \ref{eqn:unstableContribution} is the entire sensitivity, since there is no stable contribution. In Section \ref{sec:onion-results}, we observed that the integrability of the density gradient $g$ determined the existence of linear response. 
In Section \ref{sec:density-gradient}, the expression we derived for the overall sensitivity (Eq. \ref{eqn:s3}) is identical to that obtained from integration by parts of Ruelle's formula that is in Eq. \ref{eqn:unstableContribution}. Since the unstable contribution is identical to the one-dimensional case, and the stable contribution is always bounded, this same connection between the regularity of $g$ and the existence of linear response can potentially be extended to multi-dimensional systems with a one-dimensional unstable manifold (e.g. the Lorenz '63 system). For the computation of both terms of the regularized Ruelle's formula (Eq. \ref{eqn:unstableContribution}) as well as the details of the derivation above, the reader is referred to \cite{chandramoorthy-s3}. 

In practice, $g$ is computed with the following recursive formula along the trajectory (see Section 4 of \cite{chandramoorthy-s3}; \cite{sliwiak-1d} provides an intuitive explanation for the formula) 
\begin{equation}
    \label{eqn:e-iterative}
    g_{k+1} = \frac{g_k}{\alpha_k} - \frac{(\partial_{\xi} \alpha)_k}{\alpha^2_k},\;\;\;g_0 = 0,
\end{equation}
where $\alpha:=\|D\varphi\;q\|$. The subscript notation applied to a function $h$ is used to denote the composition $h_k=h\circ\varphi_{k}$. 
Note that the iterative formula in Eq. \ref{eqn:e-iterative} reduces to Eq. \ref{eqn:density-gradient-iterative} if $\varphi$ is one-dimensional. That is because, in case of 1D manifolds, the trajectory can be deformed in only one direction, which means that $q=1$, and $\alpha = |\varphi'|$. To compute $g$, we execute the recursion in Eq. \ref{eqn:e-iterative}, analogously to the 1D case described in Section \ref{sec:algorithm}. In case of higher-dimensional systems, however, the vector $q$ belongs to $E^u \neq TM$ and represents the direction of trajectory deformation as we travel along it.
It is indeed a solution to the homogeneous tangent equation $q(x_{k+1})=\alpha(x_{k}) \: D\varphi(x_k)\; q(x_k)$ with random initial conditions, and is also the only unstable Covariant Lyapunov Vector (CLV) \cite{ginelli-clv}. 

The extra mathematical difficulty in this higher-dimensional case is the evaluation of $\partial_{\xi}\alpha$, which is now an unknown, unlike in 1D systems. This term requires computing the derivative of $q$ with respect to its own direction (self-derivative), $\partial_{\xi} q$. An iterative formula, derived in \cite{chandramoorthy-clv}, can be used to approximate $\partial_{\xi} q$ along the trajectory. More details about the computation of $\partial_{\xi}\alpha$ are included in \ref{app:density-gradient}. By applying Eq. \ref{eqn:e-iterative} along a trajectory, we can generate the histogram of $|g|$ in a way analogous to the procedure described in Section \ref{sec:onion-results}.

\subsection{Probing the differentiability of statistics of the Lorenz '63 system}\label{sec:lorenz-results}
\begin{figure}
    \centering
    \includegraphics[width=1.0\textwidth]{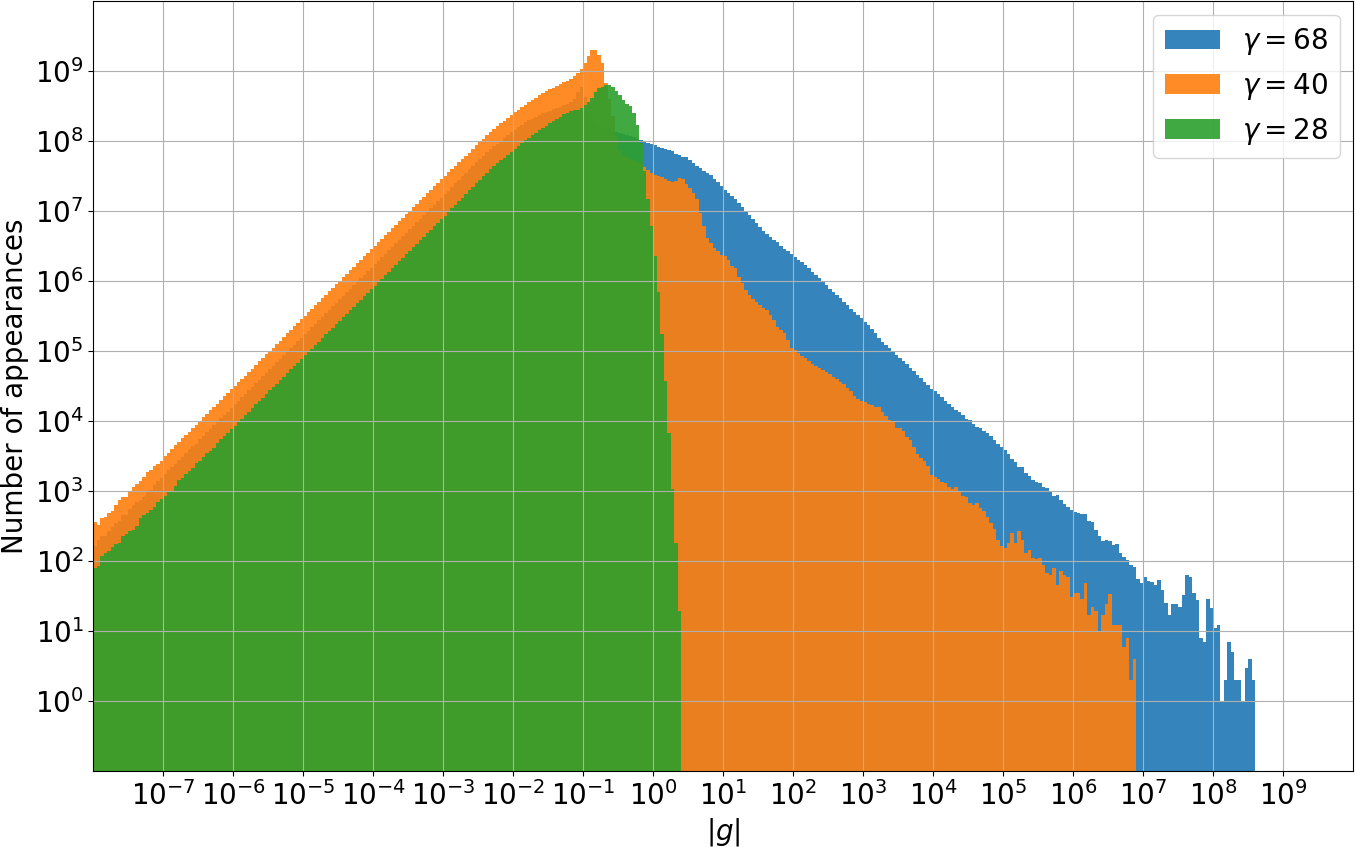}
    \caption{Distribution of the absolute value of the density gradient function generated for the Lorenz attractor at three different values of $\gamma$ using Eq. \ref{eqn:e-iterative}. To generate these histograms, we divided the x-axis from $10^{-18}$ to $10^{84}$ into $K = 2048$ bins of equal width in the logarithmic scale. For each histogram, a trajectory of the length of approximately $2.5\cdot 10^{10}$, computed by solving Eq. \ref{eqn:lorenz} with $\Delta t = 0.01$, is used.}
    \label{fig:lorenz_tail}
\end{figure}

Using the recursive relation described by Eq. \ref{eqn:e-iterative}, we generate histograms of the absolute value of the density gradient function $|g| = |\partial_{\xi}\log\rho|$ for the Lorenz '63 system. Figure \ref{fig:lorenz_tail} illustrates the distributions of $|g|$ at three different values of $\gamma$. We observe power-law behavior of the generated histograms similar to those of the onion map in Figure \ref{fig:onion_tail}. Clearly, the exponent $t$, which is introduced in Section \ref{sec:onion-results}, is much higher than 3 if $\gamma = 28$. This implies that both the expected value and variance of $|g|$ are finite, which means $g$ is square-integrable (with respect to $\rho$). The other two distributions (at $\gamma=40$ and $\gamma=68$) feature tails with exponents $t$ slightly smaller than 2, which means that $g$ may not be Lebesgue-integrable, as discussed in Section \ref{sec:onion-results}. 

Figure \ref{fig:lorenz_tail} clearly indicates that the Lebesgue-integrability threshold can be estimated to be at some $\gamma$ between 28 and 40. This result can be correlated with the regularity of the density function $\rho$ (see Figure \ref{fig:lorenz_68_density}), which apparently loses its global smoothness for 
$\gamma \leq 70$. In an extensive study of the Lorenz attractor at canonical values of $\beta$ and $\sigma$ presented in \cite{sparrow-lorenz}, it was shown that the system is quasi-hyperbolic if $\gamma\in[24.06,31]$ and non-hyperbolic if $\gamma\in[31,99.5]$. Thus, our results confirm that, in this case, the loss of hyperbolicity is also an indicator of the failure of linear response.

\begin{figure}
    \centering
    \includegraphics[width=0.48\textwidth]{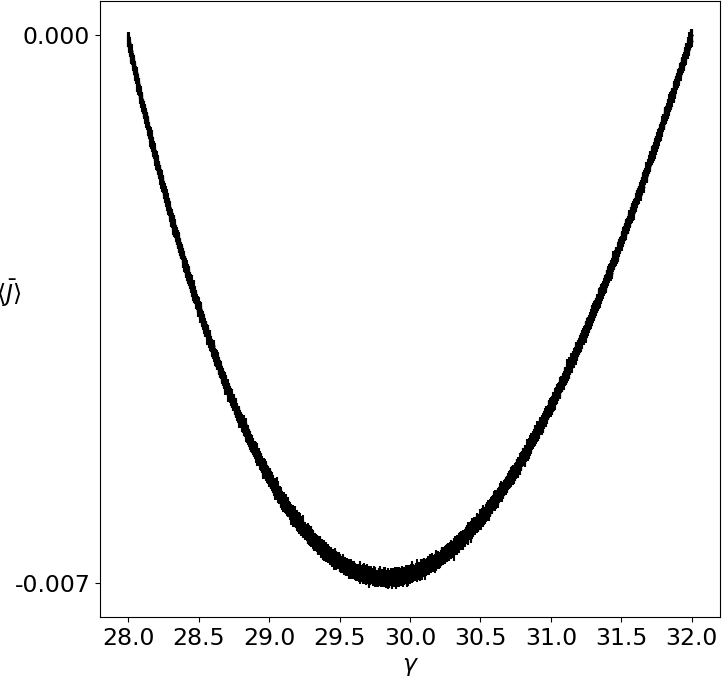}
    \includegraphics[width=0.48\textwidth]{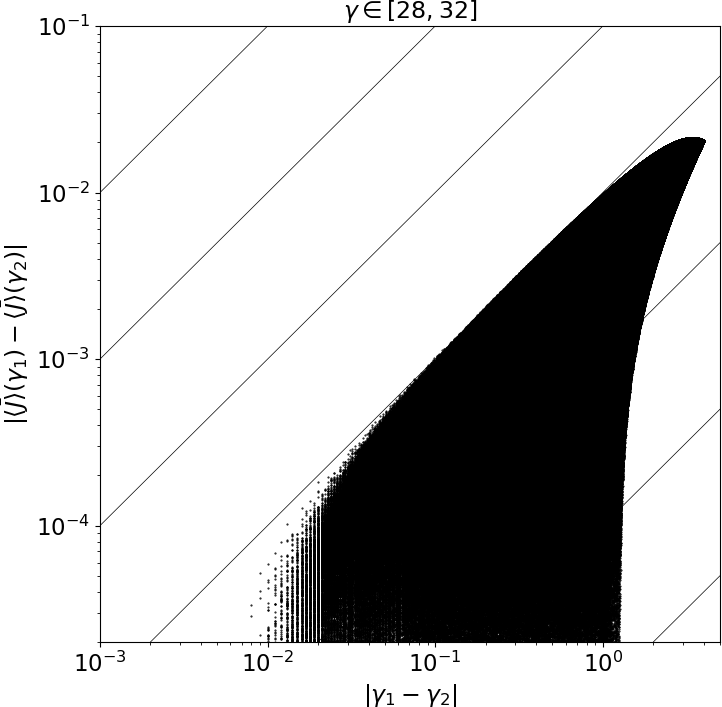}
    \\
    \vspace{5mm}
    \includegraphics[width=0.48\textwidth]{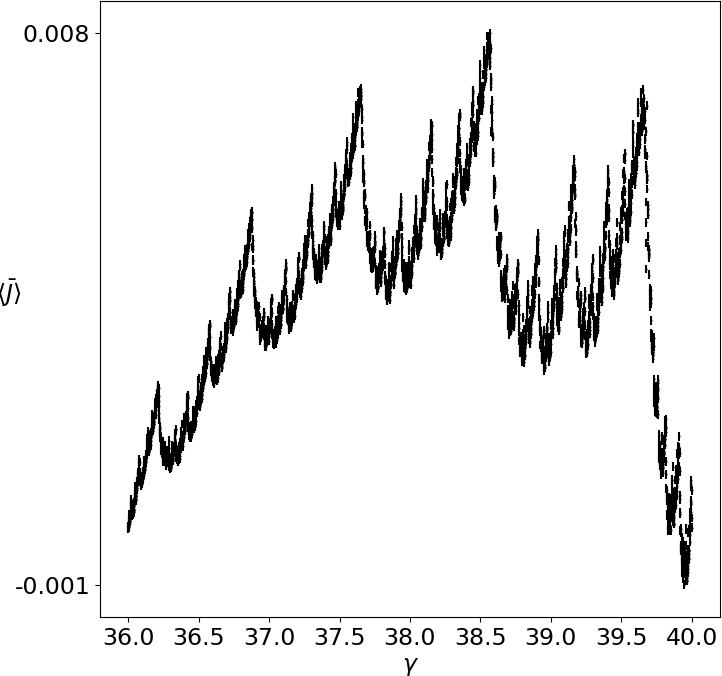}
    \includegraphics[width=0.48\textwidth]{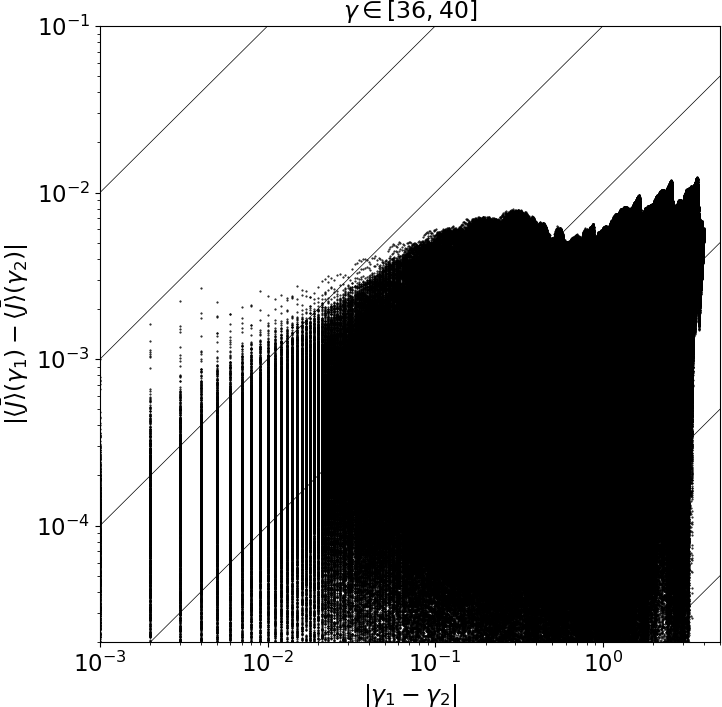}
    
    \caption{Left column: analogously to Figures \ref{fig:onion_holder1} - \ref{fig:onion_holder2}, the simulation data is the same as the data presented in Figure \ref{fig:lorenz_68_statistics}, and the quantity of interest has been modified such that the values of the long-time average at the endpoints of each interval is zero. It has been achieved by subtracting a linear function describing a straight line crossing the endpoints of the original curve, $\langle x^3\rangle - s(\gamma)$. The modified objective function has been denoted by $\bar{J}$. Each plot corresponds to a different $\gamma$ interval, which has been discretized uniformly between $\gamma_{min}$ and $\gamma_{max}$ with step size $\delta\gamma=0.001$. For each value of $\gamma$, we run 10 simulations. 
    Right column: H\"older exponent test of the statistical quantity $\langle 
    \bar{J}\rangle = \langle z\rangle -s(\gamma)$ versus parameter $\gamma$ relation plotted in Figure \ref{fig:lorenz_68_statistics}. These plots have been generated in the same fashion as those for the onion map in Figures \ref{fig:onion_holder1} - \ref{fig:onion_holder2} (see the caption of Figure \ref{fig:onion_holder1} for a detailed description), i.e., by taking the lower bound of the 3-sigma confidence interval of the data set corresponding to $[\gamma_{min},\gamma_{max}]$, obtained in 10 independent simulations. We sample the statistics every $\delta\gamma = 0.001$, which means each plot has approximately $0.5\cdot((\gamma_{max}-\gamma_{min})/\delta\gamma)^2=8\cdot 10^6$ data points. Skew solid lines represent reference lines with the slope of 1 in the logarithmic scale.}
    \label{fig:lorenz_Holder1}
\end{figure}

\begin{figure}
    \centering
    
    \includegraphics[width=0.48\textwidth]{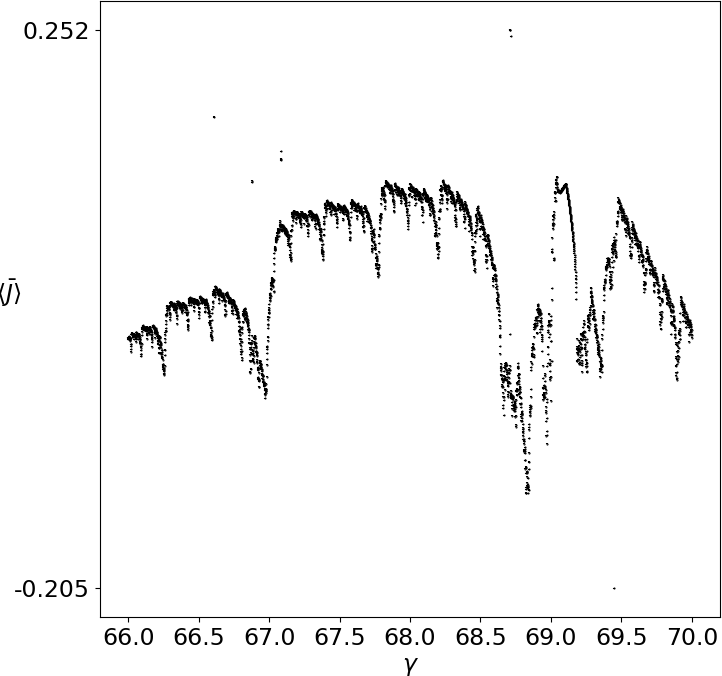}
    \includegraphics[width=0.48\textwidth]{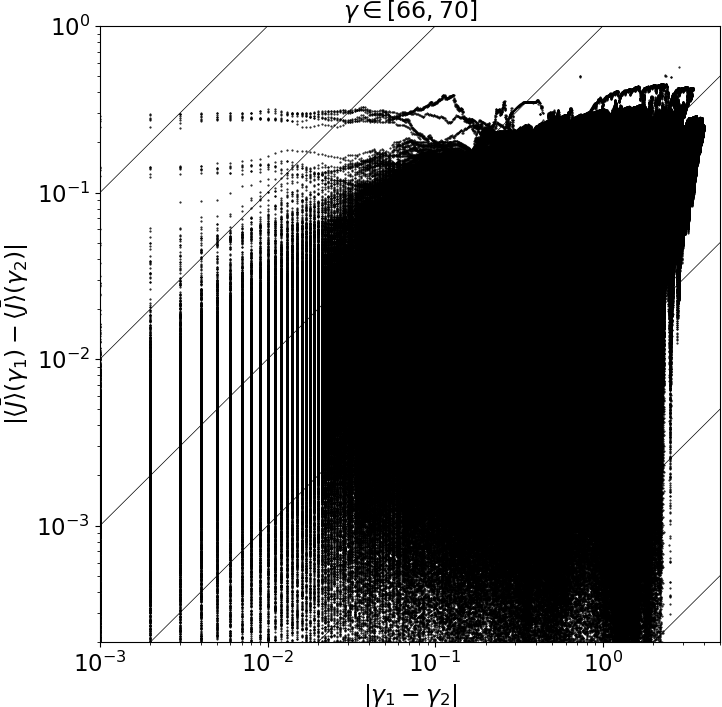}
    \caption{This figure is an extension of Figure \ref{fig:lorenz_Holder1}. All plots have been generated in the same manner as their counterparts in Figure \ref{fig:lorenz_Holder1} -- see caption therein for more details.}
    \label{fig:lorenz_Holder2}
\end{figure}

We now estimate the H\"older exponent $\mu$, as defined in Eq. \ref{eqn:holder-inequality}, for the statistics-vs-parameter relation presented in Figure \ref{fig:lorenz_68_statistics}, using the procedure described in Section \ref{sec:holder}. Figures \ref{fig:lorenz_Holder1}-\ref{fig:lorenz_Holder2} illustrate the results of the H\"older exponent numerical test generated for three different intervals of $\gamma$. These results clearly indicate the exponent $\mu$ is approximately 1 if $\gamma\in[28,32]$, implying Lipschitz-continuity of that part of the curve. The plots in the bottom row of Figure \ref{fig:lorenz_Holder1} and Figure \ref{fig:lorenz_Holder2} show $\mu$ is significantly smaller than 1, which implies the long-time average cannot be differentiable at $\gamma > 36$. Therefore, one can observe a clear correlation between the Lebesgue-integrability (with respect to $\rho$) of $g$ and smoothness of statistics, which is consistent with our numerical results of the onion map from Section \ref{sec:onion-results}.  

\section{Conclusions and future work}\label{sec:conclusions}
Statistical quantities are critical both in understanding and in applications of chaotic phenomena, such as turbulent flows. In many chaotic dynamical systems, the relation between statistical quantities and system parameters is not smooth. In this paper, we show that the existence of the parametric derivative of a statistics or long-time average (sensitivity) depends on whether the {\em density gradient} function $g$, which we define, is integrable with respect to the SRB measure. That function represents the relative rate of change of the SRB density with respect to the coordinates of the unstable manifold. The relationship between the sensitivity and $g$ is clearly reflected by the S3 formula, a closed-form expression for the sensitivity, which stems from Ruelle's linear response theory. This observation can be utilized to construct an efficient numerical procedure to assess the differentiability of statistics. The computation of the probability distribution of $|g|$ is the central part of the procedure. The probability density function of $|g|$ features a power-law behavior in case of both the systems considered in this paper: the onion map and Lorenz system. In this special case, a numerical estimate of the power law exponent is sufficient to determine the differentiability of statistics. We validate this test by numerically computing the H\"older exponent of statistics over parameter space.

The density gradient function and hence its probability distribution can be numerically generated through recursive equations along sufficiently long trajectories, solving which is the most expensive part of our procedure. These formulas require integrating the primal system, as well as first-order and second-order tangent systems in time, which in turn require the first and second derivatives of the map at each time step. 

A possible subject of future work is to generalize our algorithm to systems with higher-dimensional unstable spaces, i.e., to systems with more than one positive Lyapunov exponent. Based on the general form of the S3 formula, we believe that the major conclusion of this paper would remain the same. That is, the indicator of the smoothness of statistical quantities is still the Lebesgue-integrability of $g$. However, in systems with higher-dimensional unstable manifolds, $g$ is a vector quantity that involves derivatives with respect to all coordinates of the unstable manifold. This implies computation of derivatives of all basis vectors of the unstable space will be required. The iterative procedure for $g$, therefore, is expected to be computationally more costly.


\section*{Acknowledgments}
This work was supported by Air Force Office of Scientific Research Grant No. FA8650-19-C-2207. 

\section*{Conflict of interests}
The authors declare that they have no conflict of interests.

\section*{Supplementary materials}
In \cite{sliwiak-files}, we provide our code, including post-processing routines, that we wrote to produce Figures \ref{fig:hybrid_density}-\ref{fig:lorenz_Holder2}. In the same repository, we include most of the raw data used in the preparation of this manuscript. Some raw data files of large size were not added to the repository, however, all of them are available upon request.

\bibliographystyle{elsarticle-num-names}
\bibliography{references.bib}

\appendix

\section{Computation of density gradient in higher-dimensional systems with one positive LE}\label{app:density-gradient}

To compute the density gradient function along a trajectory, we need to differentiate $\alpha = \|D\varphi\;q\|$ with respect to $\xi$ every time step. By differentiating the Euclidean norm and using the chain rule, we can expand the derivative in the second term of the RHS of Eq. \ref{eqn:e-iterative} to the following expression,
\begin{equation}
\begin{split}
    \label{eqn:alpha-deriv}
     \partial_{\xi}\alpha = \frac{1}{\alpha}(D\varphi\;q)^T\partial_{\xi} (D\varphi\;q)= \frac{1}{\alpha}(D\varphi\;q)^T \Bigg( (D^2\varphi\cdot q)\;q+D\varphi\;\partial_{\xi}q \Bigg) := \\ \alpha (D\varphi\;q)^T\;u,
\end{split}
\end{equation}
in which $D^2\varphi$ is a third-order tensor with second derivatives of each component of $\varphi$, while the dot symbol ($\cdot$) represents the contracted tensor-vector product. In the differentiation of $\alpha$, a special type of parameterization of the unstable manifold curve $x(\xi)\in M$ is assumed, namely that $\partial_\xi x (\xi) = q (x(\xi))$. If a different parameterization was used, then the second term on the RHS of Eq. \ref{eqn:e-iterative} would have to be adjusted accordingly. Regardless of the parameterization, however, direct computation of $\|\partial_{\xi} x(\xi)\|$ is never required in the iterative process for $g$. The directional derivative $\partial_{\xi}$ measures the rate of change along the curve of the local unstable manifold, i.e., in the direction of $q$, and thus $\partial_{\xi}q$ is called the self-derivative of $q$ \cite{chandramoorthy-clv}.  Note in case of a 1D map, $q = 1$, $\partial_{\xi}q = 0$, and thus by combining Eq. \ref{eqn:e-iterative} and Eq. \ref{eqn:alpha-deriv}, we obtain the iterative formula for $g$ we derived in Section \ref{sec:density-gradient} using the Frobenius-Perron operator (compare with Eq. \ref{eqn:density-gradient-iterative}). The RHS of Eq. \ref{eqn:alpha-deriv} requires computing all possible first derivatives of the map in phase space, i.e., the $n^2$-element Jacobian $D\varphi$, and all possible second derivatives, i.e., the $n^3$-element tensor $D^2\varphi$ every time step. In case of the Lorenz '63 system, however, the tensor $D^2\varphi$ is constant, and only 4 out of its 27 elements are non-zero.\\
The final ingredient needed to evaluate $\partial_{\xi}\alpha$ is the self-derivative of $q$. Based on the covariance property of CLVs, which ensures that $(D\varphi)_k\;q_k=\alpha_k\;q_{k+1}$, and the chain rule in smooth manifolds, a recursive formula for $\partial_{\xi}q$, 
\begin{equation}
\begin{split}
    \label{eqn:iterative-2-hd}
     (\partial_{\xi}q)_{k+1} = (I-q_{k+1}\;q_{k+1}^{T})\frac{((D^2\varphi)_k\cdot q_k)\;q_k+(D\varphi)_k\;(\partial_{\xi}q)_k}{\alpha_{k}^{2}} = \\
     (I-q_{k+1}\;q_{k+1}^{T})\: u_k,
\end{split}
\end{equation}
has been derived in \cite{chandramoorthy-clv}. That study also shows Eq. \ref{eqn:iterative-2-hd} converges asymptotically at an exponential rate, regardless of the choice of the initial condition $(\partial_{\xi}q)(x_{\mathrm{init}})=(\partial_{\xi}q)_0$. Note Eq. \ref{eqn:iterative-2-hd} resembles the formula for $\partial_{\xi}\alpha$, Eq. \ref{eqn:alpha-deriv}. In fact, by applying the covariance property of CLVs, Eq. \ref{eqn:iterative-2-hd} can be rewritten as follows,
\begin{equation}
    \label{eqn:iterative-3-hd}
         (\partial_{\xi}q)_{k+1} = u_k - \frac{(\partial_{\xi}\alpha)_k}{\alpha_k^2} q_{k+1}.
\end{equation}
Therefore, the quantity $\partial_{\xi}\alpha$ is a byproduct of the recursive procedure for $\partial_{\xi}q$. We observe that the vector $u_k - (\partial_{\xi}q)_{k+1}$ is parallel to $q_{k+1}$, and the scalar $(\partial_{\xi}\alpha)_k/\alpha_k^2$ is its magnitude. 

To conclude, the computation of the density gradient function $g$ requires the iterative formula in Eq. \ref{eqn:e-iterative}. However, in case of higher-dimensional problems with one positive LE, the recursion described by Eq. \ref{eqn:iterative-2-hd} for $\partial_{\xi}q \neq 0$ must be run simultaneously, from which the scalar $\partial_{\xi}\alpha$ can be easily extracted using Eq. \ref{eqn:iterative-3-hd}.
\end{document}